\documentstyle[12pt]{article}
\newtheorem{theorem}{Theorem}[section]

\newtheorem{thm}[theorem]{Theorem}
\newtheorem{lem}[theorem]{Lemma}
\newtheorem{lemma}[theorem]{Lemma}
\newtheorem{cor}[theorem]{Corollary}

\newtheorem{example}[theorem]{Example}
\newtheorem{defn}[theorem]{Definition}
\newtheorem{definition}[theorem]{Definition}

\newcommand{\R}{{\mbox{\bf R}}}
\newcommand{\Z}{{\mbox{\bf Z}}}

\newcommand{\C}{{\mbox{\bf C}}}
\newcommand{\bS}{{\mbox{\bf S}}}
\newcommand{\D}{{\mbox{\bf D}}}

\newcommand{\M}{M}
\newcommand{\V}{V}
\newcommand{\1}{{\mbox{\bf 1}}}
\newcommand{\eps}{\epsilon}
\newcommand{\del}{\delta}
\newcommand{\rightbox}{\hfill \makebox[1 cm][r]{$\Box$}\vspace{0.5 cm}}
\newcommand{\fix}{\mbox{\rm Fix}}
\newcommand{\defined}{:=}

\newcommand{\alert}{\mbox{\bf !!!!!!!!~}}
\newcommand{\thmref}[1]{Theorem~\ref{#1}}

\newcommand{\lemref}[1]{Lemma~\ref{#1}}
\newcommand{\Equref}[1]{Equation~(\ref{#1})}
\newcommand{\eqnref}[1]{Equation~(\ref{#1})}
\newcommand{\examref}[1]{Example~\ref{#1}}
\newcommand{\corref}[1]{Corollary~\ref{#1}}

\newenvironment{pf}{{\noindent\bf Proof:}}{\rightbox}
\newenvironment{proof}{{\noindent\bf Proof:}}{\rightbox }

\input epsf

\begin{document}

\title{\bf Observing the Symmetry of Attractors}

\author{Jeffrey H.~Schenker
\thanks{Present address: Department of Mathematics, Princeton University}  
~and
James W.~Swift \thanks{Corresponding author.
E-mail: Jim.Swift@nau.edu}\\
Department of Mathematics, \\ Northern Arizona University, \\
Flagstaff, AZ 86011-5717}
\date{Received Dec. 2, 1996 by {\em Physica} {\bf D}. \\
Revised manuscript submitted June 4, 1997.}
\maketitle

\begin{abstract}
We show how the symmetry of attractors of equivariant dynamical
systems can be observed by equivariant projections of the phase space.
Equivariant projections have long been used, but they can give misleading
results if used improperly and have been considered untrustworthy.
We find conditions under which an equivariant projection
generically shows the correct symmetry of the attractor.
\\[.25cm]
\end{abstract}

\bibliographystyle{abbrv}

\section{Introduction}

Dynamical systems with symmetry often have a large-dimensional phase space,
which makes it difficult to visualize the attractors.
For example, the phase space of $6$ coupled van der Pol oscillators
is $\R^{12}$.
This paper describes how to choose a low-dimensional projection of the phase
space that shows the symmetry of a generic attractor.
In the example of $6$ coupled oscillators,
a certain 3-dimensional projection generically shows the correct symmetry of
the attractor.

Attractors of an equivariant dynamical system need not have the full symmetry 
of the system.
For example, if one tries to balance a ruler on its end, it will
inevitably fall either to the right or to the left.
Even though the system has $\Z_2$ symmetry, each of the two attractors
(right and left) has trivial symmetry.
The initial conditions determine which of the two attractors
is chosen.
This is sometimes called spontaneous symmetry breaking.

Attractors have two different types of symmetry:  {\em instantaneous} symmetry
which holds at each time step,
and {\em average} symmetry which is a symmetry of the attractor as a whole.
It is quite easy to identify the instantaneous symmetry from the coordinates
of one point on the attractor.  Therefore,
we concentrate on finding the average
symmetry of an attractor when its instantaneous symmetry is known. 
This allows lower-dimensional projections compared to an observation space
that can distinguish any possible symmetry.
Our theory also 
suggests projections to use in physical experiments, where the symmetry of the 
system is approximate.
  
Our main result is that a projection to a representation space
generically shows the proper symmetry of the attractor if the
instantaneous symmetry of the
attractor is an isotropy subgroup of the representation space.
We must say ``generically'' because it is always possible to choose a set
whose projection has too much symmetry.  We prove that an
open dense set of equivariant perturbations of the attractor project down to a
set with the correct symmetry.  This is what we mean by ``generically.''
See Theorem \ref{main} for details.

Quite often, the attractor has trivial instantaneous symmetry,
and \corref{mostUseful} applies.  In this case
a {\em faithful} representation space generically shows the correct symmetry
of the attractor.
It is easy to see from the projection if the attractor does indeed have points
with trivial isotropy.
We recommend that a faithful representation space be used to explore a
symmetric dynamical system. 
If an attractor with nontrivial instantaneous symmetry is 
observed, then choose a new projection that is
``tailored'' to the instantaneous symmetry of the attractor.

For example, in Section 3 we consider $6$ oscillators coupled in a ring with
$\D_6 \times \Z_2$ symmetry. The $\Z_2$ symmetry is included because each
term in the ordinary differential equation (ODE) is odd. 
A properly chosen equivariant projection of $\R^{12}$ onto $\R^3$
shows the correct symmetry of a generic attractor,
provided the attractor contains points with trivial symmetry
(that is, the attractor has trivial instantaneous symmetry.)
Three-dimensional space is still rather difficult to visualize.
Fortunately the symmetry can be determined by a pair of two-dimensional
projections of $\R^3$: a ``top view'' and a ``side view,'' as shown in Figure
\ref{g4.2765/6}.
Roughly speaking, the top view shows the $\D_6$ symmetry, and the 
side view is needed to distinguish between the inversion of $\Z_2$ and the
$180^\circ$ rotation in $\D_6$.

When the attractor in the $6$ oscillator system has
nontrivial instantaneous symmetry, then we can then choose
a specially tailored one- or two-dimensional projection that shows 
the symmetry of the attractor.
(See Figures \ref{1vanderPol} and \ref{delta.0}.)
On the other hand, we would need a $6$-dimensional observation space
to determine the symmetry of an attractor with any possible instantaneous
symmetry.

The recent paper ``Detecting the Symmetry of Attractors,'' by Barany et al.
\cite{Bar&al93}, introduces a method of determining the symmetry of an 
attractor
by averaging over the attractor to obtain a point in high-dimensional
representation space.  The symmetry of the attractor is the same (generically)
as the symmetry of the point.

Several papers \cite{Gol&Nic95,Tch96,Ash&Nic97} have developed
the theory of symmetry detectives,  and an alternative approach is described in 
\cite{Sch&Gra97}.  Our paper is inspired by
``Detecting the Symmetry of Attractors from Observations,'' by Ashwin \& Nicol
\cite{Ash&Nic97}. 
They separate the process into two steps: 
First the attractor is mapped, via an equivariant
map, to an isotropy equivalent observation space, then another equivariant map 
to a distinguishing representation is integrated over the image of the 
attractor.
This idea was already present in the original detectives paper
\cite{Bar&al93}, where Barany et al. discuss patterns in a square domain.

We argue that the observation step is all one needs in many cases.  
The symmetry of the attractor can easily be observed visually in a projection
of the attractor.  The averaging (detective) step converts this beautiful
picture to a point, and leads to a table of numbers.  If a number is zero, the 
attractor has a certain symmetry.  The number is never exactly zero, so 
cutoffs must be chosen.  This is not an easy task, especially since the 
appropriate cutoff depends on the dynamics of the attractor.  On the other 
hand, the human brain is excellent at pattern recognition.
The correct symmetry can be identified visually in cases where the
averaging method is inconclusive or misleading.  For example, near a
bifurcation where two conjugate attractors ``glue'' together, the eye can
recognize when the trajectory ``jumps'' from one side to another.
Then we know that the attractor has the 2-fold symmetry, although the
detective method would not identify that symmetry unless an extremely long
trajectory were averaged.

The biggest advantage of the detective method is that the process can be 
automated so that the computer can produce scans of parameter space showing the 
symmetry of the attractor(s).  This is an extremely useful tool for exploring a 
one- or two-parameter family.  Given the right software and a powerful
computer, the equivariant projection technique could be used to make a movie
of a one-parameter family of attractors.

The remainder of the paper is organized as follows:
In Section 1 we give some background in group representation
theory and describe our results.
In Section 2 we prove the theorems.
In Section 3, examples of $5$ and $6$ coupled van der Pol
oscillators demonstrate our general theory. 

\subsection{Group actions and group representations}

Throughout this paper, we let $\Gamma$ denote a finite group.
We say that a group $\Gamma$ {\em acts} on a set $M$ if there is a map
$\Gamma \times M \rightarrow M: (\gamma, x) \mapsto \gamma \cdot x$ which satisfies
$\gamma \cdot (\sigma \cdot x) = (\gamma \sigma) \cdot x$ for all $\gamma$ and
$\sigma$ in $\Gamma$, and $1 \cdot x = x$, where $1$ is the identity element in 
$\Gamma$.  (We use the ``$\cdot$" for the group action and
concatenation for group multiplication.)
If $V$ is a (real) vector space, and the action of $\Gamma$ on $V$ is linear,
we say that $V$ is a (real)
{\em representation space} for $\Gamma$ .
(We will sometimes say ``$V$ is a representation of $\Gamma$,'' even though
the homomorphism $\Gamma \rightarrow GL(V)$ induced by the action is the
representation.)
See \cite[ch. 4]{Isa94} or \cite{Ser77} for details on group actions.

\begin{defn}
A group $\Gamma$ acts {\em faithfully} on $V$ if $\gamma \cdot x = x$ for all
$x \in V$ only if $\gamma = 1$.
\end{defn}
In other words, the group action is faithful if
no nontrivial element in the group acts trivially.

\begin{defn}
Let $\Gamma$ act on $V$ and on $W$. 
Then a function $f: V \rightarrow W$ is
{\em $\Gamma$-equivariant} if $f(\gamma \cdot x) = \gamma \cdot f(x)$ for all
$\gamma \in \Gamma$.
\end{defn}
That is, the $\Gamma$ action commutes with $f$.  Note that the two ``$\cdot$"s
in the definition of equivariance are (possibly) different group actions. 
Since we know that $f$ maps $V$ to $W$, this notation is not ambiguous. 
In fact, we will often drop the ``$\cdot$'' altogether.

\begin{example}\label{D3}
Let the dihedral group $\D_3 = \langle \rho, \kappa \mid \rho^3 = \kappa^2 = 
(\rho \kappa)^2 = 1 \rangle$ act on $\R^3$ as follows:
\begin{equation}\label{D3action}
\rho \cdot (x_0, x_1, x_2) = (x_2, x_0, x_1) ~, ~~ 
\kappa \cdot (x_0, x_1, x_2) = (x_0, x_2, x_1).
\end{equation}
This action consists of permutations of the coordinates; $\rho$ is a $120^\circ$
rotation about the diagonal $x_0 = x_1 = x_2$
and $\kappa$ is a reflection through the plane
$x_1 = x_2$.  We have presented the group $\D_3$ in terms of generators and
relations, and the group action is uniquely defined by the action of the
generators.
An example of an equivariant function is
\begin{equation}
f: \R^3 \rightarrow \R^3; (x_0, x_1, x_2) \mapsto
(x_0 x_1^2 x_2^2, x_1 x_2^2 x_0^2, x_2 x_0^2 x_1^2 ),
\end{equation}
where the $\D_3$ actions on the domain and the target space are defined in 
\eqnref{D3action}.
Another equivariant function (more specifically an {\em invariant} function) is 
\begin{equation}
g: \R^3 \rightarrow \R; (x_0, x_1, x_2) \mapsto x_0 + x_1 + x_2 ,
\end{equation}
where $\D_3$ acts trivially on the target space of $g$. 
\end{example}

\subsection{Irreducible Representations}
A group action on $V$ induces a natural action on subsets $U \subseteq V$,
namely $\gamma U \defined \{ \gamma \cdot x \mid x \in U \}$.
We say that $U$ is {\em $\Gamma$-invariant} if 
$\gamma U = U$ for all $\gamma \in \Gamma$.
We are most interested in $\Gamma$-invariant sets that are also subspaces of
$V$.
\begin{definition}
A representation space $V$ is {\em $\Gamma$-irreducible} if it has no 
$\Gamma$-invariant subspaces other than $0$ and $V$.  A representation which is 
not irreducible is called {\em $\Gamma$-reducible}
\end{definition}

\begin{theorem} {\bf \em (Complete Reducibility)}\label{completeReducibility}
If $\Gamma$ is a finite group acting linearly on a finite dimensional vector
space $V$, then $V$ decomposes as 
\begin{equation}
V = V_1 \oplus V_2 \oplus \cdots \oplus V_r
\end{equation}
where each of the $V_j$ is $\Gamma$-irreducible.
\end{theorem}

A proof is found in \cite[p. 7]{Ser77}. 
The complete reducibility theorem says that
every representation is a direct sum of irreducible
representations.
Recall that $V = V_1 \oplus V_2$ 
means that every element in $V$ can be written {\em uniquely} as the sum of a 
vector in $V_1$ and a vector in $V_2$.
The direct sum is the Cartesian product of two
vector spaces which inherits vector space
addition and scalar multiplication in ``the natural'' way, {\it i.e.} 
$(v_1,v_2)
+(w_1,w_2) = (v_1+w_1,v_2+w_2)$ and $a (v_1,v_2) = (a v_1, a v_2)$.
Thus, the notation $\R^2 = \R \oplus \R$ is more explicit than
$\R^2 = \R \times \R$.

As a consequence of complete reducibility, we can choose a
basis in which the matrices of the group representation are
block-diagonal, with each block corresponding to an irreducible representation
space.
Furthermore, the basis can be chosen so that
the matrices in the group representation are orthogonal.

\begin{figure}[hptb]
\centering
\leavevmode
\epsfysize=6cm
\epsffile{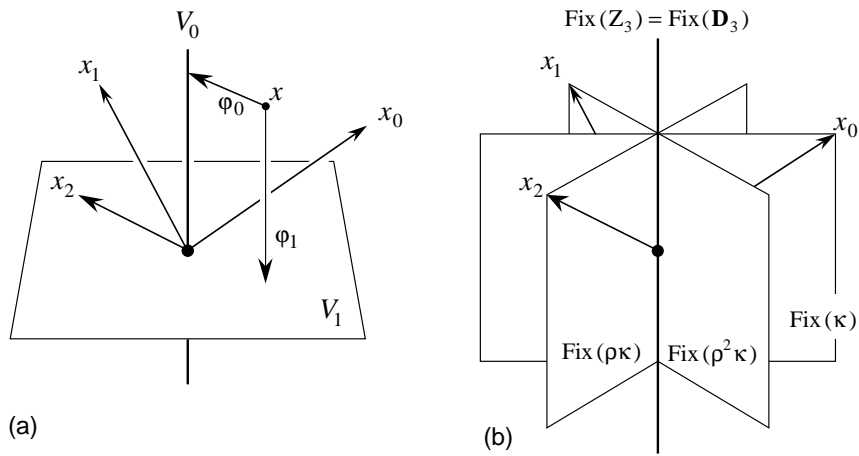}
\caption{
\label{D3fixedPointSubspaces}
The permutation action of $\D_3$ on $\R^3$.
(a) The projections onto the irreducible subspaces: $\R^3 = V_0 \oplus V_1$.
(b) The fixed point subspaces, described in Section \ref{IsotropySubgroups}.
} \end{figure}

One can always obtain an equivariant map by linear projection
onto an invariant subspace.
\begin{example}\label{D3projections}
Let $\D_3$ act on $\R^3$ as in \examref{D3}.  The decomposition into
irreducible subspaces is 
$\R^3 = V_0 \oplus V_1$, shown in
Figure \ref{D3fixedPointSubspaces}.
The projections onto 
the two irreducible subspaces are $\phi_0: \R^3 \rightarrow V_0 \cong \R$ and 
$\phi_1: \R^3 \rightarrow V_1 \cong \C \cong \R^2$, defined by
\begin{equation}
\phi_0 (x) = x_0 + x_1 + x_2.
\end{equation}
and
\begin{equation}
\phi_1 (x) =
x_0 + e^{i 2\pi/3} x_1 + e^{i 4\pi/3} x_2 .
\end{equation}
The $\D_3$ action on $V_0$ is 
trivial, and $\D_3$ acts on $V_1$ as follows:
\begin{equation}\label{D3standardAction}
\rho \cdot z = e^{i 2\pi/3} z ~, ~~~ \kappa \cdot z = \overline{z}.
\end{equation}
\end{example}

\subsection{Dynamical systems}
Let $V$ be a finite-dimensional real vector space.
There are two ways to construct a dynamical system from a map
$f:V \rightarrow V$:
\begin{enumerate}
\item The function can be used to generate a discrete system in which the
orbits,
\begin{equation}
x_0 \stackrel{f}{\mapsto} x_1 \stackrel{f}{\mapsto} x_2
  \stackrel{f}{\mapsto} x_3 \stackrel{f}{\mapsto} \cdots ~,
\end{equation}
are considered for various starting values $x_0 \in V$.
\item The function can be used to generate the vector field of a
first-order ordinary differential equation (ODE):
\begin{equation}
\dot x = f(x)
\end{equation}
In this case, the orbits are the solution curves to the ODE.
\end{enumerate}
If $f$ defines a dynamical system in either of these ways,
we will say that the dynamical system {\em stems from $f$}.

For either a discrete system or an ODE,  {\em attractors} are periodic,
quasi-periodic, or chaotic orbits to which nearby orbits `tend.' 
To define an attractor, we first define the $\omega$-limit set of a point
$x \in V$, denoted $\omega (x)$,
to be all limit points of the forward orbit through $x$.
There is no universally accepted definition of an attractor, but we take the
following:
\begin{defn}
A set $A$ is an {\em attractor} for a dynamical system if
all of the following hold:
\begin{enumerate}
\item $A$ is Liapunov stable.
(Roughly, if $x_0$ is close to $A$, then $x_n$ is close to $A$ for all $n>0$.)
\item There is an open neighborhood $U$ of $A$ such that $A$ contains
$\omega(x)$ for all $x \in U$.
\item $A$ contains a dense orbit. 
That is, $A = \omega(x)$ for some $x \in A$.
 \end{enumerate}
\end{defn}
As a consequence of this definition, an attractor is closed and
dynamically invariant.
Furthermore, an attractor has an open basin of attraction.

In this paper we are concerned with {\em equivariant dynamical systems}.
In other words, the dynamical systems stem from {\em equivariant} functions
$f: V \rightarrow V$.  We assume that the $\Gamma$ action on $V$ is faithful;
otherwise we could factor out the kernel of the group action.

\begin{lem}\label{dichotomy}
If $A$ is an attractor for a dynamical system which stems from a
$\Gamma$-equivariant function, then for each element $\gamma$ of $\Gamma$,
\begin{equation}
\gamma A = A~~\mbox{or}~~\gamma A \cap A = \emptyset.
\end{equation}
\end{lem}
\lemref{dichotomy} is proved in \cite{Cho&Gol88}. The idea is that if $A$ is
an attractor, then $\gamma A$ must also be an attractor.  Since there is
a dense orbit in $A$, and since $A$ is contained within an open basin of
attraction, $\gamma A$ is either identical to $A$ or disjoint from
$A$.  This dichotomy helps the correlation method of \cite{Sch&Gra97} work 
quite well, since the correlation
measures the extent to which ``fattened''  versions of 
$A$ and $\gamma A$ overlap.

The open basin property of attractors is crucial in the proof
\lemref{dichotomy}. 
Other definitions of attractors
do not lead to the dichotomy $\gamma A = A$ or $\gamma A \cap A = \emptyset$.
For example, attractors
with riddled basins \cite{Ale&al92} or those stuck onto an invariant subspace
\cite{Ash95} do not have open basins of attraction,
and they do not satisfy the dichotomy.  Without an open
basin of attraction an attractor is destroyed by the addition of noise.  In 
physical applications a properly defined attractor should have an open basin of 
attraction.

\lemref{dichotomy} leads to the definition
\begin{equation}
{\cal A} (V) \defined \{ A \subseteq V \mid A~\mbox{is compact
and}~\gamma A \cap A = \emptyset~\mbox{or}~\gamma A = A~\forall \gamma
\in \Gamma\}.
\end{equation}
We think of
${\cal A}(V)$ as the set of all possible attractors of an
equivariant map $f: V \rightarrow V$.

\subsection{Isotropy Subgroups and Fixed Point Subspaces}
\label{IsotropySubgroups}

Assume that $\Gamma$ acts linearly on $V$.
The {\em isotropy} of a single point $x \in V$ is 
\begin{equation}
\Sigma(x) = \{ \gamma \in \Gamma \mid \gamma x = x \} .
\end{equation}
It is easy to see that $\Sigma(x)$ is a subgroup of $\Gamma$.
We leave it as an exercise to the reader to show that $\Gamma$ acts 
faithfully on $V$ if, and only if, $V$ has a point with trivial isotropy.
(Recall that $\Gamma$ is finite.)

If $H = \Sigma(x)$ for some $x \in V$ then $H$ is
said to be an {\em isotropy subgroup} for the action of $\Gamma$ on $V$.
In general, not all subgroups of $\Gamma$ are isotropy subgroups for a given
representation space.  For example, consider the $\D_3$ action on $\R^3$
defined in \examref{D3}.  The subgroup $\Z_3 \defined \langle \rho \rangle$ is
{\em not} an isotropy subgroup, because all points which are fixed by
$\Z_3$ are on the diagonal, and these points have isotropy $\D_3$.
(When we write subgroups in terms of generators, the relations are
``inherited'' from the parent group.  Hence, $\langle \rho \rangle \leq \D_3$
is defined to be $\langle \rho \mid \rho^3 = 1 \rangle$, 
since the relation $\rho^3 = 1$ is in the definition of
$\D_3$ given in \examref{D3}.)

Ashwin \& Nicol \cite{Ash&Nic97} introduce the notion of {\em isotropy 
equivalence}, which is central to our work.
\begin{defn}\label{isoEquiv}
Two representation spaces $V_1$ and $V_2$ of a finite group $\Gamma$ are 
{\em isotropy equivalent}, written as $V_1 \sim V_2$, if $H$ is an isotropy 
subgroup for the action of $\Gamma$ on $V_1$ if and only if $H$ is an isotropy 
subgroup for the action of $\Gamma$ on $V_2$.
\end{defn}
In other words,  $V_1 \sim V_2$ when the two representation spaces have 
precisely the same isotropy subgroups.  In 
section \ref{theory} we present several lemmas concerning isotropy equivalence.
\begin{example}
Recall the $\D_3$ action on $\R^3 = V_0 \oplus V_1$ described in
Examples \ref{D3} and \ref{D3projections}.
Since $V_1$ and $\R^3$ have the same isotropy subgroups,
namely all subgroups of $\D_3$ except $\Z_3 \defined \langle \rho \rangle$,
they are isotropy equivalent: $V_1 \sim \R^3$.
\end{example}

For $A \subseteq V$ we define two types of symmetry:
\begin{enumerate}
\item The {\em instantaneous symmetry} of $A$ is
$T(A) = \{ \gamma \in \Gamma \mid \gamma x = x~\forall x \in A \} $.
This is also called the {\em pointwise symmetry}, or {\em isotropy}, of $A$.
\item The {\em average symmetry} of $A$ is
$\Sigma(A) = \{\gamma \in \Gamma \mid \gamma A = A\}$.
This is also called the {\em setwise symmetry} of $A$.
\end{enumerate}
We will usually think of $A$ as an attractor,
and the names ``instantaneous'' and ``average'' reflect this.
For singleton sets, $\Sigma(\{x\}) = T(\{x\})= \Sigma(x)$.

The instantaneous symmetry of a set is always an isotropy subgroup.
To see this, note that
\begin{equation}
T(A) = \bigcap_{x \in A} \Sigma (x).
\end{equation}
In \lemref{isotintersect} we show that the intersection of two isotropy
subgroups is an isotropy subgroup. 
Thus $T(A)$ is an isotropy subgroup.
(Since the group $\Gamma$ is finite,
we need only consider finite intersections.)

While $T(A)$ must be an isotropy subgroup, $\Sigma(A)$ can be any subgroup of
$\Gamma$, provided the group action is faithful.
To see this, let $x$ be a point
in $V$ with trivial isotropy, and let $H$ be any subgroup of $\Gamma$.  Define
$H x \defined \{ h x \mid h \in H \}$.  Then $\Sigma(H x) = H$.
For example, consider the $\D_3$ action on $\R^3$ defined in \examref{D3}.
The set $\Z_3 (0,1,2) = \{ (0,1,2), (1,2,0), (2,0,1) \}$ has average symmetry
$\Z_3$.  (Recall that $\Z_3$ is not an isotropy subgroup for this action.)

For $\Gamma$ acting linearly on $V$, we define
the {\em fixed point subspace} of $H\leq \Gamma$ to be:
\begin{equation}
\fix_{V} (H) = \{ x \in V \mid h \cdot x = x~\forall h \in H \}
\end{equation}
We summarize some facts mentioned previously and list some basic properties
of fixed point subspaces and isotropy subgroups for a linear action of
$\Gamma$ on $V$:
\begin{enumerate}
\item $\fix_{V} (H)$ is a vector space for any $H \leq \Gamma$.
\item $T(A)$ is an isotropy subgroup for any $A \subseteq V$.
\item If $H \leq G$ then $\fix_{V} (G) \subseteq \fix_{V} (H)$.
\item If $A \subseteq B$ then $T(B) \leq T(A)$.
\item $\fix_V(T(A)) = A$ if and only if $A$ is a fixed point subspace.
\item $T(\fix_V(H))=H$ if and only if $H$ is an isotropy subgroup.
\end{enumerate}
Thus, there is an inclusion-reversing isomorphism between isotropy subgroups
and fixed point subspaces.

To simplify the notation, we define
$\fix_V(\gamma) \defined \fix_V(\langle \gamma \rangle )$ for a single element
$\gamma \in \Gamma$.
Also, the
subscript ``$V$'' is often omitted.
Figure \ref{D3fixedPointSubspaces}(b) shows the fixed point spaces of the
$\D_3$ action on $\R^3$ defined in \examref{D3}. 
Since $\kappa$ interchanges $x_1$ and $x_2$ we have
$\fix(\kappa) = \{ x \in \R^3 \mid x_1 = x_2 \}$.
The diagonal in
$\R^3$ is $\fix(\Z_3) = \fix(\D_3)$. As an example of property 6, note that
$\Z_3$ is not an isotropy subgroup of $\D_3$,
and $T(\fix(\Z_3)) = \D_3 \neq \Z_3$.

Fixed point subspaces are not necessarily $\Gamma$-invariant,
but they are very important in equivariant dynamical systems
because of the following lemma:
\begin{lemma}\label{dynamInvariant}
Let $\Gamma$ act linearly on $V$ and let $f: V \rightarrow V$ be
$\Gamma$-equivariant.  If $W\subseteq V$ is a fixed point subspace,
then $f(W) \subseteq W$.
\end{lemma}
\begin{proof}
Let $H$ be the isotropy subgroup for which $W = \fix(H)$. 
Then, for each $x \in W$ and each $h \in H$,
$h f(x) = f(h x) = f(x)$.  Hence $f(x) \in W$.
\end{proof}

If a dynamical system stems from an equivariant function $f$,
\lemref{dynamInvariant} implies that fixed point spaces are
dynamically invariant. If the dynamics is an ODE, and the fixed point space
has codimension one, then the phase space is divided into dynamically
invariant sectors. 
For example, consider a 3-dimensional ODE with $\D_3$ symmetry.
Figure \ref{D3fixedPointSubspaces}(b) shows that there are 
6 wedge-shaped open regions out of which a trajectory cannot escape.  (For 
discrete dynamics in $\R^3$, an orbit {\em can} hop from one wedge to another.)

For three coupled oscillators with $\D_3$ symmetry the phase space is 
$\R^6$, one copy of $\R^3$ for position and another copy for velocity. 
Then $\fix_{{\bf R}^6}(\kappa)$
is codimension-two and there are no trapped open
regions in phase space.  In other words, the set of points in $\R^6$ with
trivial isotropy is connected.

It is now easy to see that points with isotropy $T(A)$ are dense in the
attractor $A$:  Any dense orbit must contain only points with isotropy
precisely $T(A)$, otherwise it would be ``stuck'' in a different fixed point
subspace.  However, there can be points in $A$ with larger isotropy if they
are accumulation points of a dense orbit.

On the other hand, there are sets $U$ where {\em no} points in $U$ have
isotropy $T(U)$.  For example, let $U$ be the union of the three fixed point
subspaces shown in Figure \ref{D3fixedPointSubspaces}(b). 
No points in $U$ have trivial isotropy, but $T(U) = \1$.

\subsection{Invariant Poincar\'e Sections}

Note that $\Gamma$-invariant subspaces of $V$ are not 
necessarily invariant under the dynamics. 
However, $\Gamma$-invariant subsets are important since they are candidates
for invariant Poincar\'e sections, as stated in the next theorem.
Consider, for example, an equivariant ODE with the group action shown in
Figure~\ref{D3fixedPointSubspaces}.  The
$\Gamma$-invariant subspace $V_1$ is not necessarily dynamically invariant,
but it is an invariant Poincar\'e section:
The symmetry of the attractor $A \subseteq \R^3$ is the same as the symmetry of
the intersection of $A$ and $V_1$,
provided the intersection is not empty.

\begin{thm}{\bf \em Ashwin \& Nicol} \cite[Lemma 6.2]{Ash&Nic97}.
Suppose $A$ is an attractor for $f:V \rightarrow V$, and $P\subseteq V$
satisfies (a) $\gamma P = P$ for all $\gamma \in \Gamma$ and
(b) a dense orbit in $A$ intersects $P$.
(Such a $P$ is called an {\em invariant Poincar\'{e} section}.)
Then $\Sigma(A\cap P) = \Sigma(A)$.
\end{thm}
The proof, found in \cite{Ash&Nic97}, is quite easy
considering that this important theorem seems to be new.

Note that $P$ can be any $\Gamma$-invariant set.
It can have any dimension, and it need not be a linear subspace.
Furthermore, $P$ is not really a  Poincar\'{e} section.
There is no requirement that the flow is transverse to $P$, and
a point is plotted every time the trajectory hits $P$, not just when
it crosses in one direction.

We will only consider invariant Poincar\'{e} sections $P$
which are $\Gamma$-invariant
{\em codimension-one} subspaces.  These are easily classified:  they are the
orthogonal complements to the one-dimensional representation spaces in the
decomposition of $V$ described in \thmref{completeReducibility}.
For example, with $\D_3$ acting on $\R^3$ as in
\examref{D3} and Figure \ref{D3fixedPointSubspaces},
the unique codimension-one invariant subspace
is $V_1$, which is the solution to $\phi_0(x) = x_0 + x_1 + x_2 = 0$.
For any linear group action on $V$, if $\phi$ is
the linear projection onto a one-dimensional irreducible representation of
$V$, then $\phi(x) = 0$ gives the equation for a codimension-one
$\Gamma$-invariant subspace.

\subsection{Projections which Generically Preserve Symmetry}

In this section we present what is needed to apply our theory
to the observation of symmetries of attractors. 
Those who are interested can read the theorems and proofs in Section 2. 
Others can skip Section 2 and go directly to section 3 where these ideas
are applied to systems of coupled oscillators.

A crucial observation is that an equivariant map can never decrease
the symmetry of a set:
\begin{lemma}\label{noDecrease}
Suppose that $\Gamma$ acts linearly on $M$ and $V$.  Then for any equivariant 
map $\phi: M \rightarrow V$, and any $A \subseteq M$, 
\begin{equation}
\Sigma(A) \leq \Sigma(\phi(A))
~\mbox{and} ~~ T(A) \leq T(\phi(A)).
\end{equation}
\end{lemma}
\begin{proof}
Let $\rho \in \Sigma(A)$.  Then $\rho \phi(A) = \phi(\rho A) = \phi(A)$.  Thus 
$\rho \in \Sigma (\phi (A))$. The proof for $T(A)$ is similar.
\end{proof}

Note that if the map $\phi$ is {\em invertible} we have equality in
\lemref{noDecrease},
but in general we are interested in non-invertible maps for which the domain 
is a high-dimensional phase space
and the range is a low-dimensional observation space.

The following example shows two ways in which an equivariant map can increase
the symmetry of a set.
\begin{example}\label{Z2}
Let $\Z_2 = \langle \kappa \mid \kappa^2 = 1 \rangle$ act on $\R^2$ by
$\kappa \cdot (x,y) = (-x, y)$.  
Consider the two equivariant projections $\phi_1(x,y) = x$ and 
$\phi_2(x,y) = y$. For {\em any} set $A \subseteq \R^2$, 
$\Sigma(\phi_2(A)) = T(\phi_2(A)) = \Z_2$ since the $\Gamma$ action on the
$y$-axis is trivial. 
The projection $\phi_1$ onto the 
faithful representation can also lead to the wrong symmetry.  For example let  
$A = \{ (-1, 0), (1,1) \}$.  Then $\Sigma(A) = \1$, but 
$\Sigma(\phi_1(A)) = \Sigma(\{-1, 1\}) = \Z_2$.
\end{example}

This example shows that we cannot observe the correct symmetry of 
{\em all} sets.  
The best we can hope for is that $\Sigma(\phi(A)) = \Sigma(A)$ 
when $A$ is a ``generic'' set within its symmetry class.
To make this precise, we imagine perturbing the set $A$ to $\psi(A)$,
where $\psi$ is an equivariant map from $M$ to $M$.
(If we did not require $\psi$ to be equivariant, then ``most'' perturbations
would give a set with trivial symmetry.)
The following example shows how an equivariant perturbation of $A$
can get rid of spurious symmetries in the projection.

\begin{figure}[hptb]
\centering
\leavevmode
\epsfysize=3cm
\epsffile{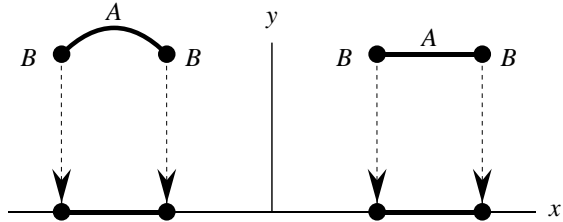}
\caption{
\label{noDichotomy}
The projection of the set $A$ onto the $x$-axis has $\Sigma = \Z_2$, even
though $A$ has trivial symmetry.  Any near-identity equivariant perturbation
of $A$ will project down to a symmetric set because the endpoints, denoted by
$B$, have $\Z_2$ symmetry.
} \end{figure}

\begin{example}
\label{perturb}
Consider the $\Z_2$ action on $\R^2$ defined in \examref{Z2}. Define the
equivariant map $\psi(x, y) = (x + \eps x \sin \pi y, y)$.
For the set $A$ defined in \examref{Z2},
the perturbed set is 
$\psi(A) = \{ (-1, 0), (1+\eps, 1) \}$, so
$\phi_1(\psi(A)) = \{ -1, 1+\eps \}$, 
which {\em does} have the same (trivial) symmetry as $A$ for $\eps \neq 0$.
\end{example}
To make progress,
we make the additional assumption 
that $A$ satisfies the dichotomy described in \lemref{dichotomy}.
Figure \ref{noDichotomy} shows what can happen without this assumption.
Let $\Z_2$ act on $\R^2$ as described in \examref{Z2}.
The set $A$ is made up of two arcs. 
The set $B$, consisting of the $4$ endpoints of the arcs,
has $\Sigma(B)=\Z_2$, even though $\Sigma(A) = \1$.
(Hence $A$ does not satisfy the dichotomy.)  
The projection $\phi_1(A)$ is two intervals, the endpoints of which are
$\phi_1(B)$, so $\Sigma(\phi_1(A)) = \Z_2 \neq \Sigma(A) = \1$. 
For any equivariant $\psi: \R^2 \rightarrow \R^2$ we have
$\Sigma(\phi_1\circ\psi(B)) = \Z_2$, since equivariant maps cannot decrease
the symmetry.
If $\psi$ is near the identity map, then $\phi_1\circ\psi(A)$
is two line segments with
endpoints $\phi_1\circ\psi(B)$.
Hence $\Sigma(\phi_1\circ\psi(A)) = \Z_2$, for an open set of $\phi$, even
though $\Sigma(A) = \1$.

It is fortunate that attractors satisfy the dichotomy of \lemref{dichotomy},
since in this case the problem shown in Figure~\ref{noDichotomy} does not 
occur. Recall that ${\cal A}(M)$ is the set of all compact subsets of $M$ 
which satisfy the dichotomy.
Ashwin \& Nicol \cite{Ash&Nic97} prove that $\Sigma(A) = \Sigma(\phi(A))$
for all $\phi$ in an open dense set of equivariant maps from $M$ to $V$,
provided that $V$ is isotropy equivalent to $M$, $A \in {\cal A}(M)$, and
points in $A$ with isotropy $T(A)$ are dense in $A$.
However, this theorem does not imply that an equivariant {\em linear} map
$\phi$ will preserve the symmetry,
and these are the ones we feel are most natural for observing the attractor.
We take a different approach, and perturb $A$ by an equivariant map
$\psi: M \rightarrow M$ as in \examref{perturb}.
Let $C^k_\Gamma(M)$ be the set of $k$-times continuously differentiable
$\Gamma$-equivariant functions from $M$ to $M$.
Our theorems give conditions on linear maps $\phi: M \rightarrow V$
which ensure that
$\Sigma(\phi \circ \psi(A)) = \Sigma(A)$ for all $\psi \in \cal{O}$, where
$\cal{O}$ is open and dense in $\C_\Gamma^k(M)$.

We now state a corollary of our main theorem, \ref{main}, which is
stated and proved in section 2. 
Consider the case where the attractor has
trivial isotropy, that is $T(A) = \1$.  Note that if any points in the
projection $\phi(A)$ have trivial isotropy,
then $T(A) = \1$ since $T(A) \leq T(\phi(A)) = \1$.

\begin{cor}\label{mostUseful}
Let $\M$ and $\V$ be finite dimensional real
representation spaces of a finite group $\Gamma$.
Let $\phi$ be a 
$\Gamma$-equivariant linear map from $M$ onto $V$.
Suppose that $A \in {\cal A}(M)$, and 
assume that $V$ is a faithful representation of $\Gamma$.
If there is a point $a \in A$ such that $T(\phi(a)) = \1$, then
\begin{equation}
T(A) = \1
~{\mbox and}~~
\Sigma(A) = \Sigma(\phi \circ \psi(A))
\end{equation}
for all $\psi \in \cal{O}$, where $\cal{O}$ is an open dense subset of
$C_{\Gamma}^k (\M)$.
\end{cor}

In practice, this means that
the symmetry of the projected attractor $\phi(A)$ is generically 
the same as the symmetry of $A$.
By ``generically," we mean that if
the symmetry of the projection is larger than the true symmetry of $A$,
then ``most'' equivariant perturbations $\psi(A)$
will project down to a set with the correct symmetry.
Since attractors of dynamical systems are quite ``generic'' sets,
we can be confident that $\phi$ shows the true symmetry if $V$ is a faithful
representation.

To observe the symmetry of the attractor with nontrivial instantaneous
symmetry, we need an observation space $V$ for which $T(A)$ is an
isotropy group, as described in \thmref{main}.
This $V$ is often smaller than a faithful representation space.
Attractors with $T(A) \neq \1$ are observed in the example of
$6$ coupled van der Pol oscillators in Section \ref{6vdP}.

\section{Theory}\label{theory}

\subsection{Isotropy Equivalence}

In this section we develop some theory regarding the notion of {\em isotropy
equivalent} representations, defined in \eqnref{isoEquiv}. 
We use the notation $V_1 \sim V_2$
to indicate isotropy equivalence when the group $\Gamma$ is understood.
Note that $\sim$ is an equivalence relation.

The following three lemmas develop several properties of isotropy subgroups.

\begin{lemma}\label{iso_group}
Let $\Gamma$ be a finite group with representation $V$.  Let 
$H \leq \Gamma$. The following are equivalent:
\begin{enumerate}
\item H is an isotropy subgroup.
\item For all $G \leq \Gamma$ such that $H \leq G$, $\fix_V (G) = \fix_V 
(H)$ if and only if G=H.
\item $\fix_V (G) \subset \fix_V(H)$ for all $G \leq \Gamma$ such that $H < G$.
\end{enumerate}
\end{lemma}
We take $\subset$ and $<$ to mean proper subset and proper subgroup 
respectively.
\\[1ex]
\begin{pf}
\begin{description}
\item[(1 $\Rightarrow$ 2)] Assume that H is an isotropy subgroup.
Let $G \leq \Gamma$ such that $H \leq G$.  Clearly if $G=H$ then 
$\fix_V (G) = \fix_V (H)$.  Suppose $\fix_V (G) = \fix_V (H)$.
Since $H$ is an isotropy subgroup, there exists $v \in \fix_V 
(H)$ such that $H=\Sigma(v)$.  Thus 
$v \in \fix_V (G)$ and since $H=\Sigma(v)$, $G \leq H$.  Thus $G=H$.

\item[(2 $\Rightarrow$ 3)] This is obvious.

\item[(3 $\Rightarrow$ 1)] Assume that $\fix_V (G) \subset \fix_V(H)$ for 
all $G \leq \Gamma$ such that $H < G$.
Let ${\cal G} = \{ G \leq \Gamma \mid H < G \}$.  Then $\fix_V(G) \subset 
\fix_V(H)$ for all $G \in {\cal G}$.  Since $\cal G$ is finite,
 $\bigcup_{G \in {\cal G}} \fix_V(G) \neq \fix_V(H)$.
Consequently there exists
\begin{equation}
v \in \fix_V(H) \setminus \bigcup_{G \in {\cal G}} \fix_V(G).
\end{equation}
Thus $H=\Sigma(v)$, so $H$ is an isotropy subgroup.
\end{description}
\end{pf}
\begin{lemma}\label{Fexist}
Let $\Gamma$ be a finite group with representation $V$.
Given $H \leq \Gamma$ there exists a unique $F \leq \Gamma$
such that $H \leq F $, $\fix_V(H)=\fix_V(F)$, and $F$ is an isotropy subgroup.
\end{lemma}
\begin{pf}
Let $F$ be the subgroup of pointwise symmetries of $\fix_V(H)$, {\it i.e.}: 
\begin{equation}
F=\{ \gamma \in \Gamma \mid \gamma v = v~~ \forall ~ v \in \fix_V(H) \}.
\end{equation}
Clearly, $H \leq F$ and $\fix_V(F) = \fix_V(H)$.
That $F$ is an isotropy subgroup follows from \lemref{iso_group}.
The uniqueness of $F$ is obvious.
\end{pf}

\begin{lemma}\label{isotintersect}
Let $H,G \leq \Gamma$ be isotropy subgroups for a 
representation $V$.  Then $H \cap G$ is an isotropy subgroup.
\end{lemma}
\begin{pf}
 Let $F \leq \Gamma$ such that $G \cap H < F$.  Then 
there exists $f \in F$ such that $f \not \in G$ or $f \not \in H$.
Without loss of generality assume $f \not \in H$.
$H$ is an isotropy subgroup, so there exists $v \in V$ such that
$H = \Sigma (v)$.  Thus $f v \neq v$. Since 
$G \cap H \leq H$, $v \in \fix_V (G \cap H)$.  Therefore, $\fix_V (F) 
\subset \fix_V (G \cap H)$.
By \lemref{iso_group}, $H \cap G$ is an isotropy 
subgroup.
\end{pf}

Note that $\Gamma$ is always an isotropy subgroup.  If the action
of $\Gamma$ on $V$ is faithful, then $\1$ is an isotropy subgroup,
otherwise the kernel of the action is an isotropy subgroup.  In fact,
\lemref{Fexist} indicates that every subgroup is contained
in a unique ``smallest''  isotropy subgroup.

Now we present two lemmas. The first of these provides a method for
showing that two representations are isotropy equivalent.  The second
allows us to reduce the size of representations by removing isotropically
equivalent representations which are ``redundant.''

\begin{lemma}\label{iso_con}
Let $\Gamma$ be a finite group with representations $V_1$ and $V_2$.
Suppose that, given $H \leq G \leq \Gamma$:
\begin{equation}
\fix_{V_1}(G) = \fix_{V_1}(H) \Longleftrightarrow
	\fix_{V_2}(G) = \fix_{V_2}(H).
\end{equation}
Then $V_1 \sim V_2$.
\end{lemma}
\begin{pf}
 This is a straight forward application of \lemref{iso_group}
\end{pf}

\begin{lemma}\label{redundant}
Let $\Gamma$ be a finite group with representations
$V'$, $V_1$, and $V_2$.  If $V_1 
\sim V_2$ then $V' \oplus V_1 \oplus V_2 \sim V' \oplus V_1$.
\end{lemma}
\newcommand{\lits}{V' \oplus V_1}
\newcommand{\bigs}{\lits \oplus V_2}
\begin{pf}
Clearly if $H$ is an isotropy subgroup for $\lits$ then $H$ 
is an isotropy subgroup for $\bigs$.  

Now let $H$ be an isotropy subgroup for 
$\bigs$. Then $H=\Sigma((v',v_1,v_2))$ for some $(v',v_1,v_2) \in \bigs$.
Since the elements of $H$ must fix $(v',v_1,v_2)$
coordinate-wise, $H \leq F \cap G_1 \cap G_2$ where $F=\Sigma(v')$, $G_1 = 
\Sigma(v_1)$, $G_2 = \Sigma(v_2)$.  However, any element of
$F \cap G_1 \cap G_2$ fixes 
$(v',v_1,v_2)$, so $F \cap G_1 \cap G_2 = H$.  

By assumption, $G_2$ is an isotropy subgroup for $V_1$.
Thus, by \lemref{isotintersect}, $G_1 \cap G_2$ is an isotropy
subgroup of $V_1$. Suppose that $G_1 \cap G_2 = \Sigma (v)$
for some $v \in V_1$.  Then $H = F \cap G_1 \cap G_2 = \Sigma (v')
\cap \Sigma (v) = \Sigma ((v',v))$.
Therefore $H$ is an isotropy subgroup for $\lits$.
\end{pf}

\subsection{$C^k_\Gamma$ Functions}

In what follows, we use the notation $\| x \|$ for the Euclidean norm of
a point $x \in \R^n$. Given a point $a \in X$, where $X$ is a metric space,
let $B(a, \del)$ denote the ball around $a$ of radius $\delta$.
We indicate the closure of a subset $B$ of a topological space by $\overline B$.

Given finite dimensional real vector spaces $\M$ and $\V$,
let $C^k (\M, \V)$ denote the set
of all $k$-times continuously differentiable functions from $\M$ to $\V$,
and define $C^k(M) \defined C^k(M,M)$.
Given $\phi \in C^k(\M,\V)$, define the standard $C^k$ norm
of $\phi$ by
\begin{equation}\label{ck}
\| \phi \|_{C^k} \defined \sup \{ \| D^{(j)} \phi (x) \| \mid  
		0 \leq j \leq k ~\mbox{and}~ x \in \M \}.
\end{equation}
When $j \geq 1$, $ D^{(j)} \phi (x) $ is the $j^{\mbox{\small{th}}}$
derivative of $\phi$ at $x$.  When $j=0$, $ D^{(j)} \phi (x)=\phi (x)$.
For $j \geq 1$, $ D^{(j)} \phi (x) $ can be defined as a $j$ linear map
from $\M$ into $\V$. In \eqnref{ck}, $\| D^{(j)} \phi (x) \|$
denotes the operator norm of $D^{(j)} \phi (x) $.

If $\M$ and $\V$ are representations of a finite group $\Gamma$
let $C_{\Gamma}^k(\M,\V)$ denote the set of $C^k$ functions
which are $\Gamma$-equivariant. 
Define $C_{\Gamma}^k(\M) = C_{\Gamma}^k(\M,\M)$.
We use the strong topology on $C^k(\M, \V)$ and $C_\Gamma^k(\M,\V)$
\cite{Hir76}.

\subsection{The Main Theorem}

We feel that equivariant linear functions are the most
natural way to visualize the phase space.
However, it is certainly not true that linear functions preserve
the symmetry of every subset of $\M$.
Our main theorem
says that if $A$ satisfies the dichotomy $\gamma A = A$ or $\gamma A \cap A =
\emptyset$, and if the linear map $\phi: M \rightarrow V$ satisfies certain
conditions, then almost all equivariant perturbations of $A$
have the correct symmetry when observed in $V$.

Before we prove \thmref{main}, recall
that given any $\phi \in C_{\Gamma}^k (\M, \V)$ and 
$A \subset \M$, $\Sigma(A) \leq \Sigma (\phi (A))$ and $T(A) \leq T(\phi (A))$
(\lemref{noDecrease}).  Recall, also, that we defined:
\begin{equation}
	{\cal A} (\M) = \{ A \subseteq \M \mid A~
	\mbox{is compact and}~\gamma A = A ~\mbox{or}~
	\gamma A \cap A = \emptyset~~\forall ~\gamma \in \Gamma \}.
\end{equation}

\begin{theorem}\label{main}
Let $\M$ and $\V$ be finite dimensional real
representation spaces of a finite group $\Gamma$.
Let $\phi$ be a
$\Gamma$-equivariant linear map from $M$ onto $V$.
Suppose that $A \in {\cal A} (\M)$ satisfies
\begin{enumerate}
  \item $T(A)$ is an isotropy subgroup for $\V$.
  \item There exists $a \in A$ such that $\Sigma(a) = T(A)$.
\end{enumerate}
Then for each positive integer $k$ there exists an open dense set ${\cal O}
\subseteq C_{\Gamma}^k (\M)$ such that
$\Sigma(\phi (\psi(a)))= \Sigma(a)$, $T(\phi(\psi(A)))=T(A)$, and
$\Sigma(\phi(\psi(A)))=\Sigma(A)$ for all $\psi \in {\cal O}$.
\end{theorem}
 
\begin{pf}
To simplify notation let ${\cal C} = C_{\Gamma}^k (\M)$. Define
\begin{equation}
{\cal O} = \{\psi \in {\cal C}
	\mid \Sigma(\phi(\psi(a)))=\Sigma(a)\ ~\mbox{and}~
	\Sigma(\phi(\psi(A)))=\Sigma(A) \}.
\end{equation}
Then for any $\psi \in {\cal O}$, $T(A) \leq T(\phi(\psi(A))) \leq
\Sigma(\phi(\psi(a))) = \Sigma(a) =T(A)$, so $T(\phi( \psi(A))) = T(A)$.
It remains to show that $\cal O$ is open and dense.

For each $\tau \in \Gamma \setminus \Sigma(a)$, let ${\cal W}_\tau
= \{\psi \in {\cal C} \mid \tau \in \Sigma(\phi(\psi(a))) \}$.
For each $\rho \in \Gamma \setminus \Sigma(A)$, let ${\cal V}_\rho
= \{\psi \in {\cal C} \mid \rho \in \Sigma(\phi(\psi(A))) \}$.
Clearly,
\begin{equation}
{\cal O} = {\cal C} \setminus  \left (
	\bigcup_{\tau \in \Gamma \setminus \Sigma(a)} {\cal W}_\tau
	\cup
	\bigcup_{\rho \in \Gamma \setminus \Sigma(A)} {\cal V}_\rho
	\right ).
\end{equation}
We show that each of the sets ${\cal W}_\tau$ and ${\cal V}_\rho$
are closed and nowhere dense.  It follows, since $\Gamma$ is finite,
that $\cal O$ is open and dense.

If $\Sigma(a) = \Gamma$, the result is trivial.  Thus suppose
that $\Sigma(a) < \Gamma$.

Let $\tau \in \Gamma \setminus \Sigma(a)$.  Define
$F : {\cal C} \rightarrow \R$ by $F(\psi) =
\|\tau \cdot \phi(\psi(a)) - \phi(\psi(a))\|$.  Clearly $F$ is
continuous and ${\cal W}_\tau = F^{-1}(0)$, so ${\cal W}_\tau$ is closed.
To show that the closed set $\cal{W}_\tau$ is nowhere dense,
it suffices to show that the complement of $\cal{W}_\tau$ in $\cal C$
is dense.
Let $\eps > 0$ and $\psi \in {\cal W}_\tau$.
We must show that there exists
$\tilde \psi \in \cal C \setminus {\cal W}_\tau$ with
$\| \psi - \tilde \psi \|_{C^k} < \eps$.
Since $T(A)$ is an isotropy subgroup for $\V$, there exists
$v \in \V$ such that $\Sigma(v)=\Sigma(a)=T(A)$.  Since $\phi$ is onto,
there exists $w \in \M$ such that $\phi(w) = v$. Let $\del > 0$
such that $\overline B(a,\del) \cap \gamma \overline B(a,\del) = \emptyset$ for
every $\gamma \in \Gamma \setminus \Sigma(a)$. 
Such a $\del$ exists because $\Gamma$ is finite and $A \in \cal A$.
Choose $\eta \in
C^k(\M,\R)$ with support $\overline B(a,\del)$ such that $\eta(a) = 1$.
Note that $\| \eta \|_{C^k} < \infty$ since $\eta$ has compact support.
Define $\tilde \psi \in {\cal C}$ by
\begin{equation}
\tilde \psi(x) \defined \psi(x) + \frac{\eps}
	{C \|\eta \|_{C^k} }
	\sum_{\gamma \in \Gamma} \eta(\gamma^{-1} x) \gamma \cdot w,
\end{equation}
where $C = \sum_{\gamma \in \Gamma} \|\gamma \cdot w \|$. Then
\begin{equation}
\phi(\tilde \psi(a)) = \phi(\psi(a)) + \frac{\eps}
	{ C \|\eta \|_{C^k}}
	| \Sigma(a) | v.
\end{equation}
So
\begin{equation}
\tau \cdot \phi(\tilde \psi(a)) - \phi ( \tilde \psi(a)) =
	\frac{\eps}{ C \|\eta \|_{C^k}}
	| \Sigma(a) | (\tau \cdot v - v) \neq 0.
\end{equation}
Thus $\tau \not \in \Sigma(\phi(\tilde \psi (a)))$, so $\tilde \psi \not \in
{\cal W}_\tau$.  But
\begin{equation}
\begin{array}{rl}
\|\psi - \tilde \psi \|_{C^k} = &
\begin{displaystyle}
\frac{\eps}
{ C \|\eta \|_{C^k}}
\left	\| \sum_{\gamma \in \Gamma} \eta(\gamma^{-1} x) \gamma \cdot w \right \|_{C^k}
\end{displaystyle}
\\
\leq &
\begin{displaystyle}
	 \frac{\eps} { C \|\eta \|_{C^k} }
	 \| \eta \|_{C^k} \sum_{\gamma \in \Gamma} \|\gamma \cdot w \| = \eps.
\end{displaystyle}
\end{array}
\end{equation}

Now we show that ${\cal V}_\rho$ is closed and nowhere dense.
If $\Sigma(A) = \Gamma$, we have nothing to prove. Therefore assume
that $\Sigma(A) < \Gamma$.

Let $\rho \in \Gamma \setminus \Sigma(A)$. Let $D(B,C)$ denote the symmetric
Hausdorff distance between compact subsets $B$ and $C$ of $\V$.  Let
$F: {\cal C} \rightarrow \R$ be defined by
$F(\psi) = D(\rho \cdot \phi(\psi (A)), \psi (A))$.
Then $F$ is continuous and ${\cal V}_\rho = F^{-1}(0)$ so ${\cal V}_\rho$
is closed. To finish the proof, we must show that
$B(\psi,\eps) \cap \left ( {\cal C} \setminus {\cal V}_\rho \right)$ is
nonempty for every $\psi \in {\cal V}_\rho$ and $\eps > 0$.
Let $\psi \in {\cal V}_\rho$ and $\eps > 0$. 
We have already shown that ${\cal C} \setminus \cup_\tau {\cal W}_\tau $
is open and dense,
in particular there exists $\hat \psi \in {\cal C}$ such that:
\begin{enumerate}
	\item $ \| \psi - \hat \psi \|_{C^k} < \eps / 2. $
	\item $ \Sigma(\phi(\hat \psi (a))) = \Sigma(a). $
\end{enumerate}
If $\rho \not \in \Sigma (\phi(\hat \psi(A)))$ then $ \hat \psi \in B(\psi,\eps) \cap
\left ( {\cal C} \setminus {\cal V}_\rho \right)$ and we are done.  Thus assume that
$\rho \in \Sigma (\phi(\hat \psi(A)))$.
In other words,
$\phi (\hat \psi (A)) = \rho \cdot \phi ( \hat \psi (A))$
even though $A \cap \rho A = \emptyset$. 
We shall perturb $\hat \psi$ so that an
extreme point of the image of $A$ in $V$ gets pushed outward from $0$,
yielding the correct symmetry of the image.
See Figure~\ref{psiHat}.

\begin{figure}[hptb]
\centering
\leavevmode
\epsfysize=4cm
\epsffile{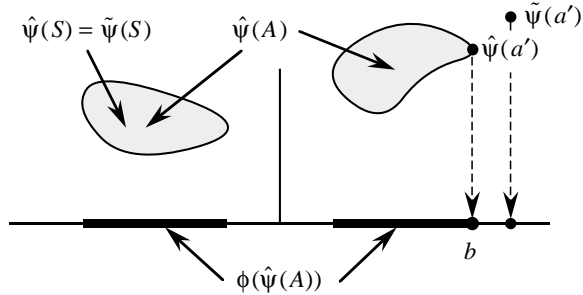}
\caption{
\label{psiHat}
The proof in the case where
$\Z_2$ acts on $\R^2$ by $(x,y) \mapsto (-x,y)$, and $\phi(x,y) =x$.
The projection of $\hat \psi(A)$ has $\Z_2$ symmetry, even
though $A$ does not.  We perturb $\hat \psi$ to $\tilde \psi$ so that
$\phi(\tilde\psi(A))$ does have the correct symmetry.
Note that $\hat\psi(A)$ need not satisfy the dichotomy that $A$ does.
} \end{figure}

Since $\phi(\hat \psi(A))$ is compact, we can
choose $t_0$ maximal such that $t_0 \phi(\hat \psi(a)) \in \phi(\hat \psi(A))$.
There exists $a' \in A$ such that
$\phi(\hat \psi(a')) = t_0 \phi(\hat \psi(a))$.
Define $b = \phi(\hat \psi (a'))$. Note that $b$ is a boundary point of
the image of $A$ in $V$, since $t b \not \in \phi(\hat\psi(A)$ for $t > 1$.
Since $\Gamma$ acts linearly and $\rho \in \Sigma(\phi \circ \hat \psi(A))$,
we conclude that $t \rho  b \not \in \phi(\hat \psi(A))$ for $t > 1$.
Further, 
$\Sigma(a) = \Sigma(a') = \Sigma(b) = T(A)$, since
$\Sigma (a') \leq \Sigma(\phi(\hat \psi(a'))) =
		\Sigma(\phi( \hat \psi(a))) = \Sigma(a) = T(A) \leq
		\Sigma(a') $.
Keep in mind that $b \neq 0$, since we are assuming that
$\Sigma(a) < \Gamma$.

Let
$S = (\phi \circ \hat \psi)^{-1}
(\{ t \rho \cdot b \mid 0 \leq t \leq 1 \}) \cap A$.
That is, $S$ contains all the points in $A$ whose image under
$\phi \circ \hat \psi$ lies in the line segment from $0$ to $\rho \cdot b$.
Since $\rho \not \in \Sigma(A)$, and $A \in {\cal A}(M)$, we know that
$\rho \cdot a' \not \in A$.
Furthermore,  $\gamma a' \not \in S$ for every
$\gamma \in \Gamma$.  Since $\Gamma$ is finite,
we can choose $\del > 0$ such that
$\gamma \overline B(a', \del) \cap S = \emptyset$ for every
$\gamma \in \Gamma$ and
$\sigma \overline B(a', \del) \cap \overline B(a', \del) = \emptyset$
for every $\sigma \not \in \Sigma(a')$.  Choose a non-negative function
$\eta \in C^k(\M, \R)$ with support
$\overline B(a',\del)$ and $\eta(a') = 1$. 
Let $\chi$ denote the characteristic function of the group orbit
$\{ \gamma \overline B(a',\del) \mid \gamma \in \Gamma\}$.
Define $\tilde \psi \in {\cal C}$ by
\begin{equation}
\tilde \psi (x) \defined \left ( 1 + \frac{\eps}
		{2^{k+1} | \Gamma | \|\eta \|_{C^k} \|\hat \psi \chi \|_{C^k}}
		\sum_{\gamma \in \Gamma} \eta(\gamma x) \right ) \hat \psi(x).
\end{equation}
Note that $\phi(\tilde \psi (x)) = \phi(\hat \psi(x))$ for every $x \in S$.
Also,
$\phi (\tilde \psi (x)) \propto \phi(\hat \psi(x))$ for every $x \in A$,
with a constant of proportionality greater or equal to $1$.
In particular,
\begin{equation}
\phi(\tilde \psi (a')) = \left ( 1 + \frac{\eps}
		{2^{k+1} | \Gamma | \|\eta \|_{C^k} \|\hat \psi \chi\|_{C^k}}
		|\Sigma(a') | \right ) \phi(\hat \psi(a')) = t b.
\end{equation}
with $t > 1$.
Given $x \in A$, $\phi(\tilde \psi(x))  = t \rho \cdot b$
only if $0 \leq t \leq 1$ in which case $x \in S$.   Thus,
$\rho \cdot \phi(\tilde\psi(a')) \not \in \phi(\tilde \psi(A))$.  Therefore
$\tilde \psi \not \in {\cal V}_\rho$.  Furthermore,
\begin{equation}
\begin{array}{rl}
\| \tilde \psi - \psi \|_{C^k} \leq &
\|\tilde \psi - \hat \psi \|_{C^k} + \| \hat \psi  - \psi \|_{C^k} \\
	<  &
\begin{displaystyle}
   \frac{\eps}
          {2^{k+1} | \Gamma | \|\eta \|_{C^k} \|\hat \psi \chi\|_{C^k}}
		\sum_{\gamma \in \Gamma} \|\eta(\gamma x) \hat\psi(x) \chi(x) \|_{C^k}
    + \eps / 2
\end{displaystyle}
\\
	\leq &
\begin{displaystyle}
    \frac{\eps}
          {2^{k+1} | \Gamma | \|\eta \|_{C^k} \|\hat \psi \chi\|_{C^k}}
		2^k | \Gamma | \|\eta \|_{C^k} \|\hat \psi \chi \|_{C^k} + \eps / 2
   = \eps.
\end{displaystyle}
\end{array}
\end{equation}	
In the second line we used the fact that
$\eta(\gamma x) \hat \psi(x) =\eta(\gamma x) \hat \psi(x) \chi (x)$ for all
$\gamma \in \Gamma$. 
This complication is necessary because $\| \hat \psi \|_{C^k}$
might be infinite.
In the third line we used the fact, easily demonstrated with Leibnitz' rule,
that $\| fg \|_{C^k} \leq 2^k \|f\|_{C^k} \|g\|_{C^k}$ for any
$f \in C^k(M,\R)$ and $g \in C^k(M,M)$.

Thus ${\cal V}_\rho$ is closed and nowhere dense.
\end{pf}

Note that the two-step procedure of perturbing $\psi$ to $\hat \psi$ and then
to $\tilde \psi$ is necessary when $\psi(x) = 0$ for all $x \in M$. 

Corollary \ref{mostUseful} follows from Theorem \ref{main} if we restrict to
$T(A) = \1$, and recall that a representation space of a finite group
is faithful if it has points with trivial isotropy.

\section{Oscillators coupled in a ring}

As an example of an equivariant dynamical system, we
consider $n \geq 3$ coupled van der Pol oscillators:
\begin{equation}\label{vanderPol}
	\begin{array}{l}
    \begin{displaystyle}
			\dot{x_j} = y_j - \delta (y_{j-1} + y_{j+1} )
    \end{displaystyle} \\
    \begin{displaystyle}
			\dot{y_j}  =  -x_j + y_j (\alpha - x_j^2 )
 			+ \beta ( y_{j-1} + y_{j+1}) + \gamma ( y_{j-2} + y_{j+2})
			+ \delta ( x_{j-1} + x_{j+1})
    \end{displaystyle}
  \end{array}
\end{equation}
where $j \in \{ 0, 1, \ldots , n-1 \}$ is defined modulo $n$. We can think of 
the oscillators coupled in a ring with coupling to their nearest neighbors 
(the $\beta$ and $\delta$ terms) and their next-to-nearest neighbors (the 
$\gamma$ term).  
We investigated \eqnref{vanderPol} using {\em dstool} driven by
a fourth order Runge-Kutta algorithm with fixed time step $h = 0.01$.

The equations have the symmetry $\Gamma = \D_n \times \Z_2$.
The symmetry of the regular $n$-gon, $\D_n$, is due to 
the geometric symmetry of the ring. 
We write the $n$-fold rotation in $\D_n$ as $\rho$, and the reflection as 
$\kappa$.  Thus
$\D_n \defined \langle \rho, \kappa :
\rho^n = \kappa^2 =  (\rho \kappa)^2 = 1 \rangle $.
(Note that \cite{Kro&Ste95} and \cite{Sch&Gra97} use $\zeta$ for the rotation.)
The symmetry group acts on $\R^{2n} = \R^n \oplus \R^n$ as two copies of the 
same representation, one for the $x$ coordinates and one for the $y$ 
coordinates. For simplicity we show the action on the $x$ coordinates only:
\begin{eqnarray}
\rho \cdot (x_0, x_1, \ldots, x_{n-1})
& = & (x_1, x_2, \ldots , x_0) \\
\kappa \cdot (x_0, x_1, \ldots, x_{n-1})
& = & (x_0, x_{n-1}, x_{n-2}, \ldots, x_1) 
\end{eqnarray}
Note that $\kappa$ acts trivially when $n=2$.  This is why we
assume $n \geq 3$ throughout this paper.

The $\Z_2$ symmetry in $\Gamma = \D_n \times \Z_2$ is present because
every term in  
\Equref{vanderPol} has odd degree. The equations are symmetric under
the inversion through the origin in $\R^{2n}$:
\begin{equation}
\label{inversion}
\sigma \cdot (x, y) = -(x,y).
\end{equation}
Clearly $\sigma$ commutes with all elements in $\D_n$.
It is very common for oscillator models to have this inversion symmetry.

\subsection{Irreducible Representations of $\D_n$ and $\D_n \times \Z_2$}
When we refer to $\D_n$ as the symmetries of a regular $n$-gon, we are
implicitly thinking of the action on the plane, which is most simply written
in terms of complex coordinates $z \in \C \cong \R^2$:
\begin{equation}
\label{V1action}
\rho \cdot z = e^{i 2 \pi/n} z ~, ~~~~ \kappa \cdot z = \overline{z}
\end{equation}
The complex coordinate is used to denote the 
2-dimensional {\em real} representation $\Gamma \rightarrow O(2)$;
\begin{equation}
\label{realrep}
\rho \mapsto 
\left (
\begin{array}{rr}
\cos(2 \pi / n) & - \sin(2 \pi / n) \\
\sin(2 \pi / n) & \cos(2 \pi /n )
\end{array} \right )
~, ~~~~ \kappa \mapsto
\left (
\begin{array}{rr}
1 & 0 \\
0 & -1 
\end{array} \right )
\end{equation}
\eqnref{V1action} does not give a 1-dimensional {\em complex}
representation $\Gamma \rightarrow U(1)$ because 
of the complex conjugate $\overline{z}$.

The group action in \eqnref{V1action} can be generalized to give
an irreducible representation space $V_m$ for
every integer $m$,
where $\D_n$ acts on $z_m \in V_m$ as follows:
\begin{equation}\label{2Dreps}
\rho \cdot z_m = \left(e^{i 2 \pi/n}\right)^m z_m ~, 
~~ \kappa \cdot z_m = \overline{z_m}
\end{equation}
The $V_m$ are also irreducible representation spaces of
$\D_n \times \Z_2$, where the generator of $\Z_2$ acts as
\begin{equation}
\sigma \cdot z_m = - z_m
\end{equation}
Note that
\eqnref{V1action} is the group action on $V_1$.
The representation space $V_m$ is the same as $V_{n+m}$ and 
it is conjugate to $V_{n-m}$.
Hence, we only need consider $m$ in
the set $0 \le m \le n/2$.
When $0 < m < n/2$, $V_m \cong \C$ is a (real) 2-dimensional representation
space, whereas $V_0\cong \R$ and $V_{n/2}\cong \R$ (if $n$ is even)
are one-dimensional representation spaces.
The rotation $\rho \in \D_n$ acts as a rotation by $2 \pi m/n$ radians in $V_m$.

\eqnref{2Dreps} does not describe all of the irreducible representations
of $\D_n$.  There is a one-dimensional 
irreducible representations in which $\rho \mapsto 1$ and $\kappa \mapsto -1$.  If 
$n$ is even, there is another one-dimensional 
irreducible representations in which $\rho \mapsto -1$ and $\kappa \mapsto -1$.

We now introduce linear equivariant functions of the phase space:
Let $(x, y) \in \R^{2n}$, and for each integer $m$ define
$v_m \in V_m$ and $w_m \in W_m$ as follows:
\begin{equation}\label{DnChange}
v_m = \phi_m(x,y) \defined
\sum_{j=0}^{n-1} x_j \left(e^{i 2 \pi/n}\right)^{mj}~,~~~ 
w_m = \psi_m(x,y) \defined
\sum_{j=0}^{n-1} y_j \left(e^{i 2 \pi/n}\right)^{mj}~.
\end{equation}

\begin{figure}[hptb]
\centering
\leavevmode
\epsfxsize=13cm
\epsffile{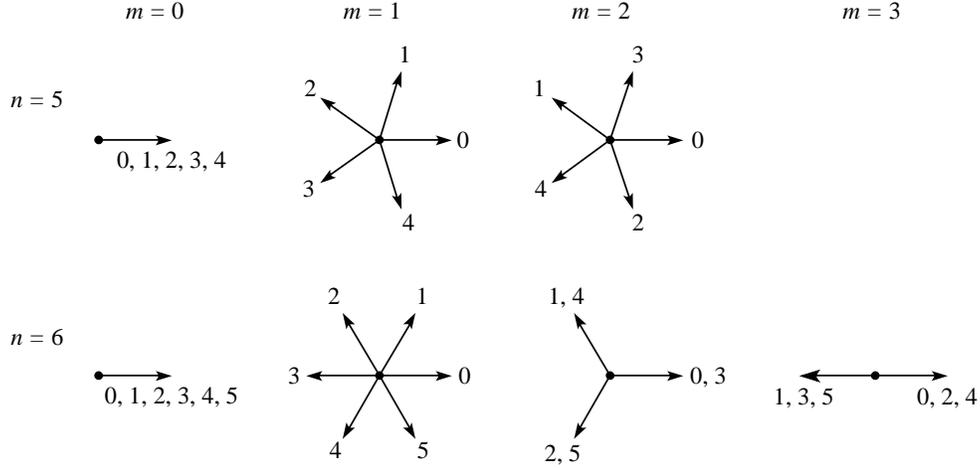}
\caption{
\label{rootsOfUnity}
This shows the various roots of unity used to define $v_m = \phi_m(x,y)$
and $w_m = \psi_m(x,y)$ for $n = 5$ and $n=6$ oscillators.
} \end{figure}

Figure~\ref{rootsOfUnity} gives a graphical interpretation of this sum:
The coordinates $x_j$ or $y_j$ are multiplied by various roots of unity
in the complex plane.

The functions $\phi_m$ and $\psi_m$ defined in \eqnref{DnChange}
are equivariant linear projections onto the irreducible representation spaces
$V_m$ and $W_m$, respectively, of $\D_n$.  (These are also irreducible 
representation spaces of $\D_n \times \Z_2$.)
Note that $\phi_m$ generalizes the projections $\phi_0$ and $\phi_1$ defined
in \examref{D3projections} for $n=3$.
For a given $m$, $W_m$ is simply a copy of $V_m$, and we could have written
$v_m, w_m \in V_m$.  Instead we write $v_m \in V_m$, and $w_m \in W_m$ to
distinguish the two copies of the same representation space.
Note that $\phi_m = \phi_{m+n}$ and $\phi_{n-m} = \overline{\phi_m}$. 
We obtain a linear 
change of basis by considering only $0 \le m \le \lfloor n/2 \rfloor$ (the 
floor of $n/2$).
In fact, the decomposition of the phase space described in 
\lemref{completeReducibility} is
\begin{equation}
\label{istopicDecomp}
M = \R^n \oplus \R^n =
V_0 \oplus V_1 \oplus \cdots \oplus V_{\lfloor n/2 \rfloor}
\oplus
W_0 \oplus W_1 \oplus \cdots \oplus W_{\lfloor n/2 \rfloor}
\end{equation}
The group action is block-diagonalized in the $(v, w)$ coordinates:
\begin{equation}
\begin{array}{ll}
\rho v_m  =   \left(e^{i 2 \pi/n}\right)^m v_m ~,~~~ &
	\rho w_m = \left(e^{i 2 \pi/n}\right)^m w_m \\
\kappa v_m =  \overline{v_m} ~,~~~ & \kappa w_m = \overline{ w_m} \\
\sigma v_m  =  - v_m ~,~~~ & \sigma w_m = - w_m .
\end{array}
\end{equation}
Note that only the irreducible representations of
$\D_n$ described in \eqnref{2Dreps} are present in the
decomposition of the phase space.

Recall that a linear projection onto a {\em faithful} representation space
generically preserves symmetries of attractors with trivial isotropy
(see \corref{mostUseful}).
\begin{lemma}
The projection $\phi_1: \R^{2n} \rightarrow V_1$ is a linear projection onto a 
faithful representation space of $\D_n$. For the group $\D_n \times \Z_2$, 
$V_1$ is a faithful representation space if $n$ is odd, whereas $V_0 \oplus 
V_1$ is the smallest faithful representation space if $n$ is even.
\end{lemma}
\begin{proof}
This is straightforward. 
When the group is $\D_n \times \Z_2$ and $n$ is even, 
$\rho^{n/2} \sigma v_1 = v_1$,
so the action on $V_1$ is not faithful.
\end{proof} 

In order to observe any possible symmetry, including cases with nontrivial 
instantaneous symmetry, we need a representation space that is isotropy 
equivalent to the phase space $\R^{2n}$.
\begin{lemma}
\label{gcdLemma}
The representations $V_q$ and $V_r$ of $\D_n$ or $\D_n \times \Z_2$ are 
isomorphic, and hence isotropy equivalent, if and only if $\gcd(q,n) = 
\gcd(r,n)$. (Recall that $\gcd(q,n)$ is the
greatest common divisor of $q$ and $n$.)
\end{lemma}
\begin{proof}
The root of unity in \eqnref{DnChange} can be replaced by any other primitive 
root of unity, namely $e^{i 2 \pi p/n}$ where the $\gcd(p, n)=1$.  
(Note that $p$ and $n$ are relatively prime precisely when $\gcd(p,n)=1$.)
We call the projection $\tilde{\phi}$ when using $e^{i 2 \pi p/n}$,
and $\phi$ when using $e^{i 2 \pi /n}$.  Hence
\begin{equation}
\tilde{\phi}_m (x, y) =  
\sum_{j=0}^{n-1} x_j \left(e^{i 2 \pi p/n}\right)^{mj}
= \phi_{pm} (x, y)
\end{equation}
But $\phi_m$ and $\tilde{\phi}_m$ are isomorphic since they are induced by the 
outer automorphism of $\D_n$ defined by $\rho \mapsto \rho^p$, $\kappa \mapsto 
\kappa$.
Therefore $V_m \cong V_{pm}$ provided $p$ is relatively prime to $n$.
(We use ``$\cong$'' to denote isomorphism, and ``$\sim$'' to denote isotropy
equivalence. Note, however that $V_m$ and $V_{pm}$ are not conjugate
representations, and they have different characters.)
The isomorphic representations correspond to the cycles of the permutation 
$m \mapsto pm$ modulo $n$, where $\gcd(p,n)=1$.  The
$\gcd(m,n)$ is constant for every element of a cycle, and different for every 
cycle.  (For example, the cycle decomposition of $m \mapsto 3m$ modulo 10 is 
$(0)(1397)(2684)(5)$, and the greatest common divisors are $0$, $1$, $2$, and 
$5$, respectively.)  Hence, $V_q \cong V_r$ if $\gcd(q,n) = \gcd(r,n)$.  
Conversely, if $\gcd(q,n) \neq \gcd(r,n)$ the representations cannot be 
isotropy equivalent because the general point in $V_q$ has isotropy 
$\langle \rho^{\gcd(q,n)}\rangle$ whereas the general point in $V_r$ has 
isotropy $\langle \rho^{\gcd(r,n)}\rangle$.  
The proof is unchanged if the group is $\D_n \times \Z_2$ rather than $\D_n$. 
\end{proof}

For example, for $n=10$, $V_1 \cong V_3$ and $V_2 \cong V_4$.
\begin{theorem}
For the permutation action of $\D_n$ on $\R^{2n}$,
the smallest representation space 
isotropy equivalent to $\R^{2n}$ is the direct sum of all $V_m$ where
$m$ divides $n$ and
$m < n$:
\begin{equation}
\R^{2n} \sim \bigoplus_{m | n, ~  m < n} V_m .
\end{equation}
For $\D_n \times \Z_2$, the smallest 
representation space isotropy equivalent to $\R^{2n}$ also includes $V_0$:
\begin{equation}
\R^{2n} \sim V_0 \oplus \bigoplus_{m | n, ~ m < n} V_m .
\end{equation}
\end{theorem}
\begin{proof}
\lemref{redundant} states that that $V^\prime \oplus V \oplus W \sim 
V^\prime \oplus V$ if $V \sim W$.  
Note that $V_m$ and $W_m$ (with the same $m$) are conjugate and therefore 
isotropy equivalent, since there is an invertible linear change of
coordinates which interchanges $x_j$ and $y_j$ for all $j$.
Thus, we can ``throw away'' the $W_m$ representations in the decomposition 
\eqnref{istopicDecomp}.  
Lemma~\ref{gcdLemma} implies that we only need $V_m$ where $m$ is a 
factor of $n$. 
(This includes $V_n$, which is the same as $V_0$.)
For the group $\D_n$, $V_0$ is the trivial representation and 
can be eliminated.  Finally, we cannot get rid of $V_0$ for $\D_n \times \Z_2$, 
since the point $(1, 0) \in V_0 \oplus V_1$ has isotropy $\D_n$,
but no point in 
$V_1$ has this isotropy ($0 \in V_0$ has isotropy $\D_n \times \Z_2$).
\end{proof}

For example, for $n=10$ coupled oscillators in \eqnref{vanderPol}, the smallest 
observation space which is isotropy equivalent to the phase space
$\R^{20}$ is
$\V_0 \oplus \V_1 \oplus V_2 \oplus V_5$, which is 6 dimensional
($1 + 2 + 2 + 1 = 6$).

\subsection{Five coupled van der Pol oscillators}

The case of $n=5$ van der Pol oscillators is typical of systems with $\D_n 
\times \Z_2$ symmetry when $n$ is prime.  The 
representations $V_1$, $V_2$, $W_1$ and $W_2$ are all isotropy equivalent and 
faithful. 
(We remark that $V_1$ and $V_2$ are {\em not} lattice equivalent, contrary
to the claim in \cite{Bar&al93}.)  
While $V_0 \oplus V_1$ is the smallest representation space that is isotropy 
equivalent to the 10-dimensional phase space, we only need to look at $V_0$ 
when the attractor projects to a point at the origin in $V_1$.  

Here is an algorithm to determine the symmetry of the attractor, assuming that 
$A$ is a ``generic'' set:
\begin{enumerate}
\item
If $\phi_1(A) = 0$, then $T(A) \geq \D_5$.
In words, if the $V_1$ projection of the attractor is a point at 
the origin, then either the oscillators are in-phase ($T(A) = \D_5$), 
or the attractor is a point at the origin in $\R^{10}$
($T(A) = \D_5 \times \Z_2$). Consider then the projection onto $V_0$:

(a)  If $\phi_0(A) = 0$, then $T(A) = \Sigma(A) = \D_5 \times \Z_2$.

(b)  If $\phi_0(A) = -\phi_0(A) \neq 0$, then $T(A) = \D_5$ and $\Sigma(A) = 
\D_5 \times \Z_2$. 

(c)  If $\phi_0(A) \neq - \phi_0(A)$, then $T(A) = \Sigma(A) = \D_5$.

In practice, we would look at the $(v_0, w_0)$ plane rather than the one 
dimensional projection onto $V_0$.  In case (b) the figure
would have 2-fold rotational symmetry in $V_0 \oplus W_0$,
whereas in case (c) it would not.
\item
If $\phi_1(A) \neq 0$, then the symmetry of $A$ is the symmetry of the 
projection onto  $V_1$.  The instantaneous symmetry is $T(A) \cong \D_1$ if the 
attractor lies in a line of reflection, and $T(A) = \1$ otherwise. 
Note that $\D_5 \times \Z_2 \cong \D_{10}$, and $\sigma \in \Sigma(A)$ 
precisely when $\phi_1(A) = - \phi_1(A)$. (Recall that $\sigma$ is the
generator of $\Z_2$.)
\end{enumerate}

\begin{figure}[hptb]
\centering
\leavevmode
\epsfxsize=14.5cm
\epsffile{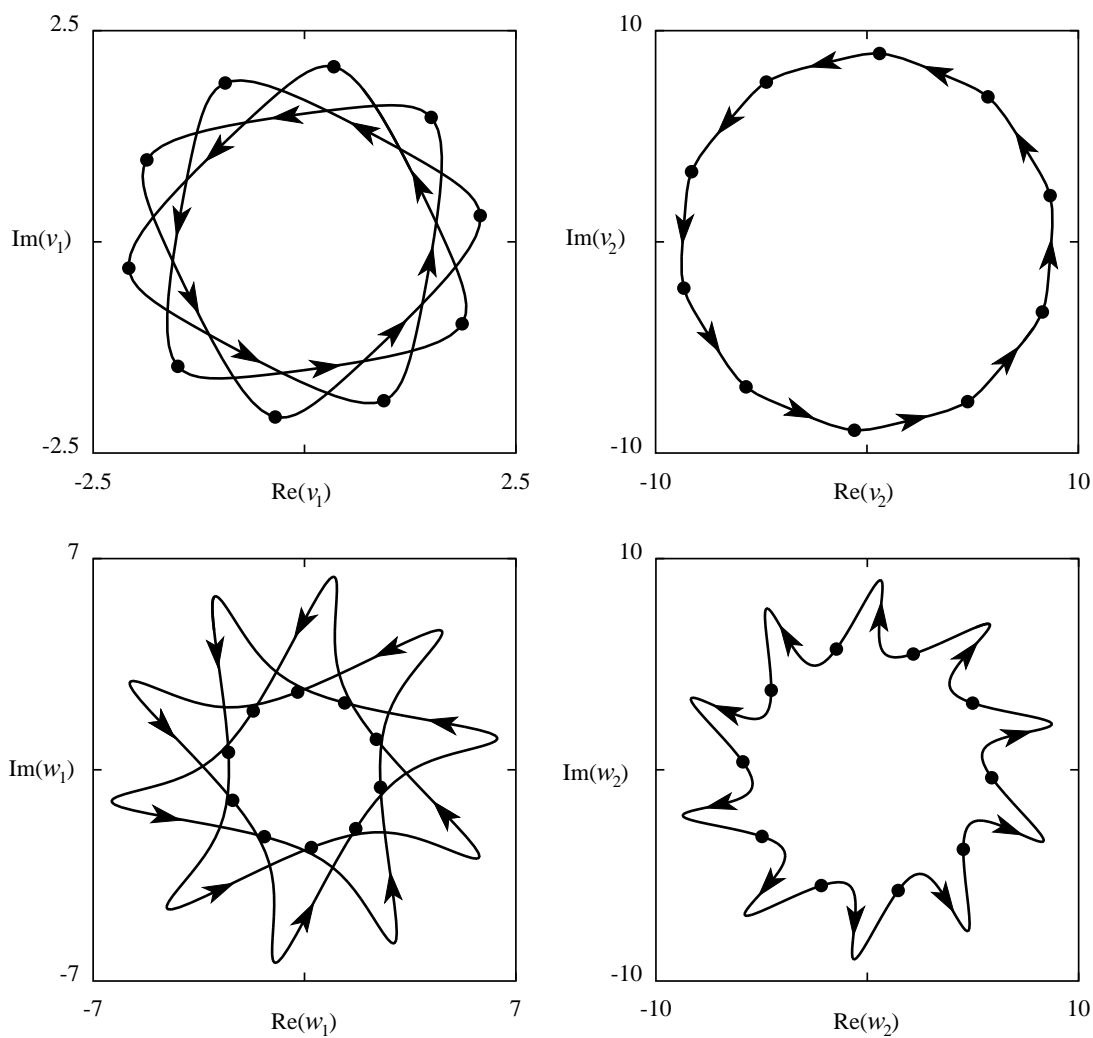}
\caption{
\label{5osc}
A periodic attractor for $5$ coupled van der Pol oscillators in the
projections onto $V_1$, $V_2$, $W_1$, and $W_2$.
The dots are where the trajectory intersects the invariant Poincar\'e section
$w_0 = 0$.
Note that the
symmetry is that same in each projection: only one is needed to observe the
symmetry of the attractor.
The Runge-Kutta step size is $h=0.01$ throughout this paper.
} \end{figure}

Figure~\ref{5osc} shows the attracting periodic orbit of \eqnref{vanderPol} for 
$n=5$, $\alpha = 2$, $\beta = -0.5$, and $\gamma = \delta = 0$.  The negative 
$\beta$ coupling pushes adjacent oscillators out of phase.  We show the 
projections onto $V_1$ and the other faithful representations.
The large dots in Figure~\ref{5osc} are an invariant 
Poincar\'{e} section obtained by plotting a point whenever $w_0 = 0$.

As expected, the Poincar\'{e} section has the same symmetry as the full
trajectory, and the symmetry in each of the projections is the same:
Since there 
are points on the attractor with trivial symmetry, $T(A) = \1$.  The 
10-fold rotational symmetry of the attractor indicates that 
$\Sigma(A) = \Z_5 \times \Z_2 \cong \Z_{10}$.  
The reflections in $\D_5 \times \Z_2 \cong \D_{10}$
act on the plane as reflections across lines at the angles
$0$, $\pi / 10$, $2 \pi / 10$, $\ldots$, $9 \pi / 10$.
The attractor does not have any of these reflectional symmetries.
The attractor in each Poincar\'e section consists of 10 equally spaced points
on a circle, and hence it has reflectional symmetry about 10 lines
of reflection, but these are not at the ``correct'' angles and hence they
are not symmetries of the attractor.
(The limit cycles in the $V_1$ and $V_2$ projections appear to have these
``spurious'' symmetries as well, perhaps because the
$x_j(t)$-waveforms are nearly sinusoidal.)

By considering the time order that the points on the Poincar\'e section are
visited, we can determine that
the attractor has the 
spatio-temporal symmetry $x_j(t+T/10) = -x_{j+1}(t)$, where $T$ is the 
period of the limit cycle.  (Spatio-temporal symmetries are not included in
$\Sigma(A)$.)

Note that the projections of the limit cycle cross the lines of reflection 
symmetry (e.g. $\rm{Im}(v_1) = 0$)
even though every point in the attractor has trivial isotropy.  These 
are ``non-generic'' points that project to points with the wrong symmetry.
This illustrates how the symmetry of the attractor can be too big.
For example, consider the Poincar\'e section in $V_1$. 
The 10 points of the attractor will shift around as a
parameter is varied, and at isolated parameter values one of the points will
just happen to fall on the line $\rm{Im}(v_1) = 0$, and we would conclude
wrongly that $\Sigma(A) = \D_5 \times \Z_2$.
Fortunately, such coincidences are very unlikely with chaotic attractors.

\subsection{Six coupled van der Pol oscillators}
\label{6vdP}
As our second example we choose 6 oscillators because 6 is rich in factors.  
The smallest faithful representation of $\D_6 \times \Z_2$ is 3-dimensional: 
$V_0 \oplus V_1$, and the smallest representation isotropy equivalent
to $\R^{12}$ is 6-dimensional: $V_0 \oplus V_1 \oplus V_2 \oplus V_3$.

First we explored \eqnref{vanderPol} with $\delta = 0$.  Although we did not do 
an exhaustive study, we never saw quasiperiodic or chaotic dynamics.
Either the attractor was a single point at the origin, with symmetry $T(A) =
\Sigma(A) = \D_6 \times \Z_2$, or it was a limit cycle with 
nontrivial instantaneous symmetry, as shown in Table 1. 
The crucial hypothesis in Theorem \ref{main} is that the instantaneous symmetry
$T(A)$ should be an isotropy subgroup of the observation space $V$.
This is enough to proceed, however a short theoretical discussion describing
the rest of Table 1 is illuminating.

Assume that $T(A)$ is nontrivial (that is, $T(A) \neq \1$).
To simplify notation, let
$T \defined T(A)$.
Recall that $A \subseteq \fix_M(T)$.  
Hence $\Sigma(\fix_M (T))$ is the largest
possible average symmetry that $A$ can have. 
It is an interesting fact that $\Sigma(\fix_M(T)) = N_\Gamma(T)$,
the normalizer of $T$ in $\Gamma$, which is defined for the abstract group 
without reference to the representation.
We define
$\Gamma_T \defined N_\Gamma(T)/T$, which is the largest possible symmetry
of $A$ modulo the instantaneous symmetries.
(Beware that \cite{Gol&Nic95} and \cite{Ash&Nic97} use $T'$ where we write
$\Gamma_T$.)
Thus, the ``effective'' phase space is $\fix_M(T)$, and the ``effective''
symmetry group is $\Gamma_T$.

\begin{table}[hptb]
$$
\begin{array}{ccccc}
T \defined T(A)
& V
& \Gamma_T
& \fix_{{\bf R}^6}(T)
&  \Sigma(A)
\\
\hline
\langle \rho , \kappa \rangle \cong \D_6 
& V_0 
& \Z_2 
& \{( a, a, a, a, a, a )\} 
& \D_6 \times \Z_2 
\\
\langle\rho\sigma,\kappa\rangle \cong \D_6 
& V_3 
& \Z_2 
& \{( a,-a, a,-a, a,-a )\} 
& \D_6\times\Z_2 
\\
\langle \rho^3\rangle \cong \Z_2
& V_2 
& \D_3 \times \Z_2
& \{( a, b, c, a, b, c )\} 
& \langle \rho,\sigma\rangle \cong \Z_6 \times \Z_2 
\\
\langle\rho^3\sigma\rangle \cong \Z_2
& V_1 
& \D_3 \times \Z_2 
& \{(a,-b,c,-a,b,-c )\} 
&\langle\rho,\sigma\rangle \cong \Z_6 \times \Z_2 
\end{array}
$$
\caption{
Observed symmetry of attractors of \eqnref{vanderPol} with $n=6$ and
$\delta = 0$.
The first column is the instantaneous symmetry $T$, and
$V$ is the representation space for which $T$ is an isotropy subgroup.
The effective symmetry group and phase space are $\Gamma_T$ and
$\fix_{{\bf R}^6}(T) \oplus \fix_{{\bf R}^6}(T)$.
The last column show the average symmetry $\Sigma(A)$, which 
was always observed to be the same for a given $T$.
}\end{table}

\begin{figure}[hptb]
\centering
\leavevmode
\epsfxsize=6.5cm
\epsffile{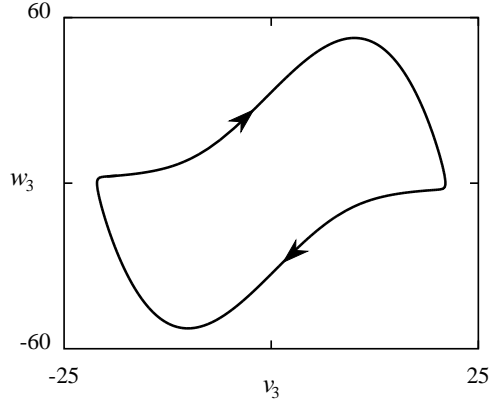}
\caption{
\label{1vanderPol}
A periodic attractor of \eqnref{vanderPol} with $n = 6$, $\alpha = 1$,
$\beta = -1$, and $\gamma = \delta = 0$. 
The symmetry is described in the second row of Table 1:
$T = \langle \rho \sigma, \kappa \rangle$.
} \end{figure}

Figure~\ref{1vanderPol} shows the projection onto $V_3 \oplus W_3$ for a limit
cycle with $T(A) = \langle \rho \sigma, \kappa \rangle$. 
The symmetry can be deduced by the projection onto $V_3$ alone,
but since $V_3$ is one-dimensional we can draw pictures in $V_3 \oplus W_3$.
Each oscillator is $\pi$ out of phase with its nearest neighbors.
Within the effective phase space  $\fix_{{\bf R}^{12}}(T)\cong \R^2$,
\eqnref{vanderPol} reduces to
a {\em single} van der Pol oscillator with effective symmetry
$\Gamma_T \cong \Z_2$ 
\begin{equation}
	\begin{array}{l}
    \begin{displaystyle}
			\dot{x} = y + 2 \delta y
    \end{displaystyle} \\
    \begin{displaystyle}
			\dot{y}  =  -x + y ((\alpha-2\beta+2\gamma-2\delta) - x^2 )
    \end{displaystyle}
  \end{array}
\end{equation}
where $v_3 = 6x$ and $w_3 = 6y$ are the coordinates in $V_3 \oplus W_3$.
The attractor has full symmetry, $\Sigma(A) = \D_6 \times \Z_2$,
since the limit cycle has inversion symmetry.

\begin{figure}[hptb]
\centering
\leavevmode
\epsfxsize=7cm
\epsffile{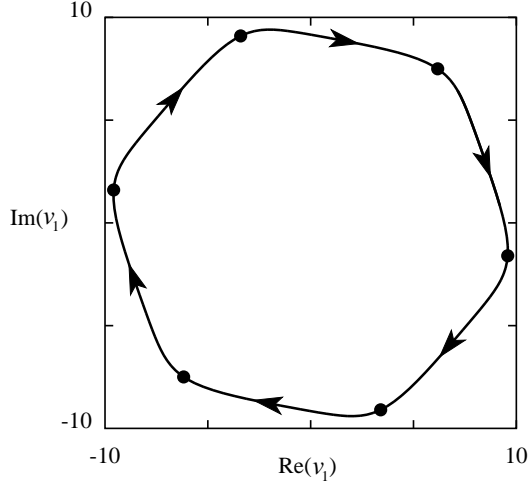}
\caption{
\label{delta.0}
A periodic attractor for $6$ coupled van der Pol oscillators in the $V_1$ 
projection. 
The dots are where the trajectory intersects the invariant Poincar\'e section
$w_0 = 0$.
The parameters in \eqnref{vanderPol} are $\alpha = 1$,
$\beta = 0.75$, $\gamma = -0.5$, and $\delta = 0$.
The symmetry is described in the last row of Table 1:
$T = \langle \rho^3 \sigma \rangle$.
} \end{figure}

The cases where $T \cong \Z_2$ reduce to \eqnref{vanderPol} with $n=3$,
so there are effectively 3 van der Pol oscillators with
$\D_3 \times \Z_2$ symmetry.
Figure~\ref{delta.0} shows the $V_1$ projection of a periodic attractor
with $T = \langle \rho^3 \sigma \rangle$, the last row in Table 1. 
We can determine $T$ in two ways:
First, the $x_i$ coordinates of a single point on the
trajectory match the pattern of $\fix_{{\bf R}^6}(T)$ in Table 1.
(The $y_i$ coordinates also match this pattern.)
Alternatively, the signature of this instantaneous symmetry is that the
projections onto $V_0$ and $V_2$ are $0$,
while there are non-zero projections onto $V_1$ and $V_3$.
Once the instantaneous symmetry is known, $\Sigma$ can be determined by the 
$V_1$ projection alone.
We can think of $\Gamma_T$ as $\langle \rho^2, \kappa, \sigma \rangle$.
The elements in $\Gamma_T$ that give the symmetry
observed in the $V_1$ projection are 
$\langle \rho^2, \sigma \rangle \cong \Z_3 \times \Z_2 \cong \Z_6$.
Thus, the full symmetry is
$\Sigma (A) = \langle \rho^2, \sigma, \rho^3 \sigma \rangle =
\langle \rho, \sigma \rangle$. 

When $\delta \neq 0$ more 
exotic symmetries and quasiperiodic or chaotic dynamics are common.  
As an example, consider the one-parameter family with 
$\alpha = 1$, $\beta = 0$, $-2 \geq \gamma \geq -5$, and $\delta = -0.1$.  
The instantaneous symmetry is $T(A) = \1$ throughout. 
The average symmetry and dynamics are summarized in Table 2. 

\begin{table}
\centering
\leavevmode
\epsfxsize=\linewidth
\epsffile{figures/table2.epsi}
\caption{
The observed dynamics and symmetry
in \eqnref{vanderPol} 
for the family with $n=6$,
$\alpha = 1$, $\beta = 0$, $\delta = -0.1$ and $\gamma$ the parameter.
The dynamics are either periodic (P), quasiperiodic (Q), or chaotic (C).
The averaged symmetries are
$\Sigma (A) = \D_6 \times \Z_2$,
$\Z_6 \times \Z_2 = \langle \rho, \sigma \rangle$,
$\Z_6 = \langle \rho \rangle$,
$\Z_3 \times \Z_2 = \langle \rho^2, \sigma \rangle$,
or
$\Z_3 = \langle \rho^2 \rangle$.
We changed $\delta$ by steps of $0.1$ to produce the table,
and found bistability only at $\gamma = -2.3$.  
}\end{table}

Unlike the cases with 
nontrivial instantaneous symmetry, we are forced to use a 3-dimensional 
projection, $V_0 \oplus V_1$, which is the smallest faithful representation
space.
We used the invariant Poincar\'{e} section defined by $w_0 = 0$.
We plotted two different 2-dimensional projections of $V_0 \oplus V_1$ (the
``top'' view and the ``side'' view) on the computer screen.
The top view is the 
$V_1$ plane, and the side view shows
the real part of $v_1$ horizontally and $v_0$ vertically.
We used three colors to signify $v_0 < -\eps$,
$| v_0| < \eps $, and $v_0 > \eps$, for an appropriate small $\eps$. 
Color figures of each type of attractor listed in Table 2 can be found
at the web site {\tt http://odin.math.nau.edu/$\sim$jws/}~~.

For this article, we show black and white figures of selected attractors.
(The color would be needed to distinguish between
$\Sigma(A) = \langle \rho^3, \kappa, \sigma \rangle$ and
$\Sigma(A) = \langle \rho^3, \kappa \sigma \rangle$, but we never
observed either of these two symmetries.)

The figures we show in this article concern the bifurcation that occurs between
$\gamma = -4.2$ and $\gamma = -4.3$.  From Table 2 we see that the symmetry
changes from $\Z_3$ to $\Z_6 \times \Z_2$. 
The number of conjugate attractors, which is $| \Gamma | / |\Sigma |$ in
general, goes from $8$ to $2$.
Does this change of symmetry occur in 2 distinct bifurcations?
Much to our surprise, the answer is no.

\begin{figure}[hptb]
  \centering
  \leavevmode
  \epsfysize=10cm
  \epsffile{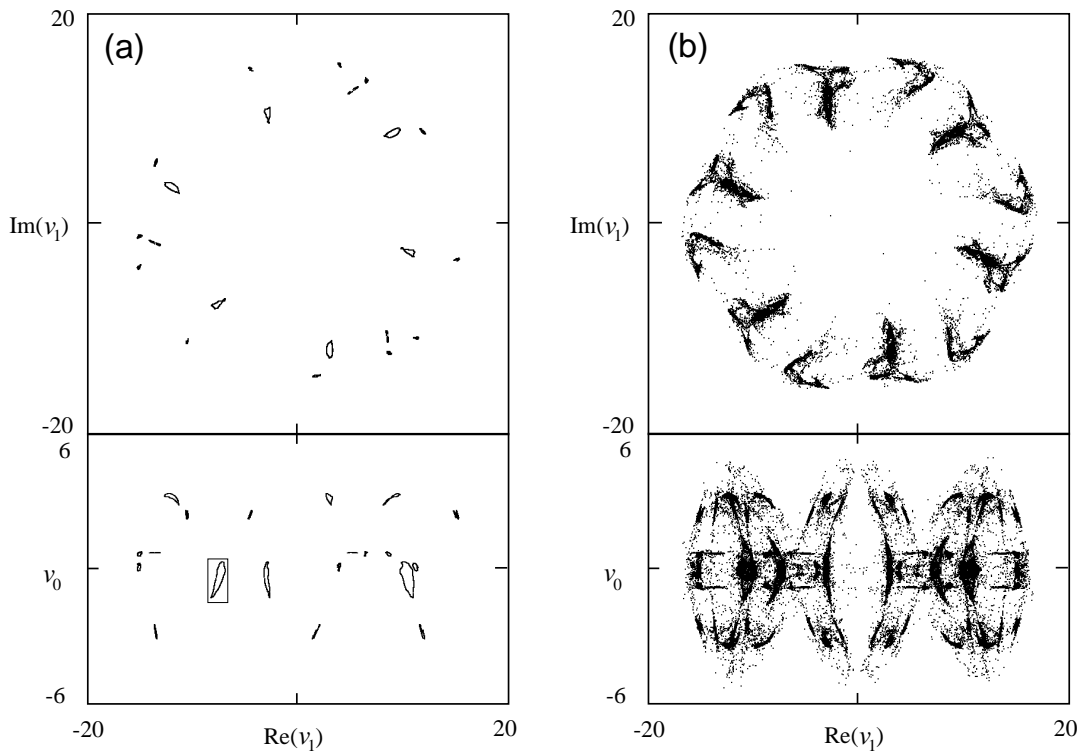}
\caption{
\label{g4.2765/6}
Two projections of the periodic attractor when (a) $\gamma = -4.2765$
and (b) $\gamma = -4.2766$.
The other parameters are $\alpha = 1$, $\beta = 0$, and $\delta = -0.1$.
The upper ``top'' view shows the projection onto $V_1$.
In the lower ``side'' view, the horizontal
coordinate is the real part of $v_1$ and the vertical coordinate is $v_0$.
Points on the invariant Poincar\'e section are plotted whenever $w_0 = 0$.
There are 50,000 points in(a) and 200,000 points in (b). 
}
\end{figure}

As $\gamma$ decreases, the periodic orbit at $\gamma = -4.2$ has a Hopf
bifurcation to an invariant torus, and then the torus becomes crinkled, giving
a presumably chaotic attractor at $\gamma = -4.2765$,
shown in Figure~\ref{g4.2765/6}(a).
From the $V_1$ projection (the top view) there are two possibilities:
$\Sigma = \langle \rho^2 \rangle$ or
$\Sigma = \langle \rho^2, \rho^3 \sigma \rangle$.
(Recall that $\rho^3 \sigma$ acts trivially on $V_1$.)
The side projection in Figure~\ref{g4.2765/6}(a)
shows that $\rho^3 \sigma \notin \Sigma$. 
Hence the symmetry is $\Sigma = \langle \rho^2 \rangle = \Z_3$,
unchanged from the symmetry of the periodic orbit at $\gamma = -4.2$ listed in
Table 2.

Figure~\ref{g4.2765/6}(b) shows the chaotic attractor at $\gamma = -4.2766$. 
Here the average symmetry is
$\Sigma(A) = \langle \rho, \sigma \rangle = \Z_6 \times \Z_2$. 
It is clear from the side view that $\langle \rho^3, \sigma \rangle$
is a subgroup of $\Sigma(A)$.
The attractor $A$ in Figure \ref{g4.2765/6}(a), along with the conjugate
attractors $\sigma A$, $\rho^3 A$, and $\rho^3 \sigma A$, merge into the
single attractor in Figure \ref{g4.2765/6}(b) at a single bifurcation.
There is no hysteresis, except for the fact that in going from (b) to (a) one
of the four conjugate attractors is chosen randomly.
The trajectory of the attractor in 
Figure~\ref{g4.2765/6}(b) spends several thousand iterates
near one the four ``ghost'' attractor left over from figure (a).
Then it escapes and stays far from the ghosts for
a few hundred iterates until it ``falls'' into
any of the ghost attractors.
Eventually, the trajectory visits all four ghosts.  The symmetry of the
attractor thus changes from $\langle \rho^2 \rangle$ to
$\langle \rho^2, \rho^3, \sigma \rangle = \langle \rho, \sigma \rangle$
at a single bifurcation.

\begin{figure}[hptb]
  \centering
  \leavevmode
  \epsfysize=6cm
  \epsffile{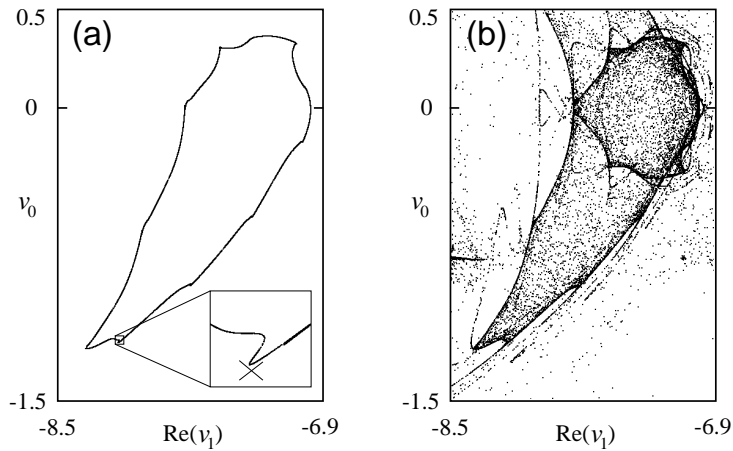}
\caption{
\label{sic}
Blow-up of the two attractors in Figure \ref{g4.2765/6},
showing the symmetry increasing
crisis where $4$ attractors merge into one.  One million points on the
invariant Poincar\'e section are computed at each parameter value.
The inset in (a) suggests that the dynamics is chaotic.
A ``ghost'' attractor, and part of its conjugate under $\rho^3 \sigma$,
are seen in (b).
}
\end{figure}

This can be explained as a {\em symmetry increasing crisis}.
Figure \ref{sic} shows a blow-up of the boxed area in the side view of
Figure \ref{g4.2765/6}(a).
The folds in the attractor shown in Figure \ref{sic}(a) indicate that
the invariant circle has broken down in a standard bifurcation from
quasiperiodicity to chaos.
We can infer the existence of a saddle-type
periodic orbit of the Poincar\'e map,
indicated by an ``x'' in the inset of Figure \ref{sic}(a).
Somewhere between $\gamma = -4.2765$ and $\gamma = -4.2766$ the attractor
collides with the periodic orbit in a homoclinic bifurcation (the symmetry
increasing crisis).
At the bifurcation,
one side of the one-dimensional unstable manifold is homoclinic, while
the other side of the unstable manifold is in the
basin of attraction of all four ghost attractors.
This is what causes the random hopping between the four ghost
attractors.

This symmetry increasing crisis is different from
the collision of two attractors discussed in \cite{Cho&Gol88}
and \cite{Del&Hei95}.
In a collision, the size of the attractor doubles.
In the crisis we observed, the attractor more than quadruples in size.
A similar bifurcation with two ghost attractors
was described as an explosion in \cite{Del&al92}.
Crises of attractors, when no symmetry is present, are
discussed in \cite{Gre&al83}.
Typically, a crisis is accompanied by hysteresis,
but in a symmetry increasing crisis there is no hysteresis.

\section{Conclusion}

We have given simple and easily verifiable conditions to ensure that the 
projection of an attractor in an equivariant dynamical system has the same 
symmetry (generically) as the attractor.  If the attractor has trivial 
isotropy, then the projection to a faithful representation space is sufficient. 
This method is not always practical.  For example, consider dynamical systems 
with $\bS_n$ permutation symmetry acting on $\R^n$.  Then the smallest faithful 
representation is $(n-1)$-dimensional. 
For a group like $\bS_{10}$ acting on $\R^{10}$ the projection 
technique is hopeless.  On the other hand, symmetry detectives or any other 
method are also bound to fail.

For group actions where
the smallest faithful representation is 3-dimensional (or higher),
progress can be made by looking at various 2-dimensional projections, 
using color, or by 
rotating a 3-dimensional projection of the object in real time on the computer 
screen. 
For example, with $\D_n \times \Z_2$ acting on $\R^n$, the smallest faithful
representation is three-dimensional. 
Figure \ref{g4.2765/6} demonstrates how two different
2-dimensional projections, the top view and the side view,
show the symmetry of the attractor.

\paragraph{Acknowledgement}
This research was supported by the NSF through the Research Experience for
Undergraduates (REU) program.
We would like to thank Peter Ashwin, Matt Nicol, Mike Field, Mike Falk,
Michael Dellnitz, John Hagood, Farhad Jafari, and Steve Fromm 
for helpful discussions.

\bibliography{paper}

\end{document}